\documentclass[aps,preprintnumbers, superscriptaddress, nofootinbibt,twocolumn]{revtex4}
\usepackage{amsmath}
\usepackage{eurosym}
\usepackage{amssymb}

\pdfoutput=1

\usepackage{graphicx}

\usepackage{color}

\usepackage{braket}

\def\be{\begin{equation}}
\def\ee{\end{equation}}
\def\bea{\begin{eqnarray}}
\def\eea{\end{eqnarray}}

\begin{document}

\title{Minimum length uncertainty relations in the presence of dark energy}
\author{Matthew J. Lake}
\email{matthew5@mail.sysu.edu.cn/mjlake@ntu.edu.sg}
\affiliation{School of Physics, Sun Yat-Sen University, Guangzhou 510275, China}
\affiliation{School of Physical and Mathematical Sciences, Nanyang Technological University, 637371 Singapore, Singapore}


\begin{abstract}

We introduce a dark energy-modified minimum length uncertainty relation (DE-MLUR) or dark energy uncertainty principle (DE-UP) for short. 
The new relation is structurally similar to the MLUR introduced by K{\' a}rolyh{\' a}zy (1968), and reproduced by Ng and van Dam (1994) using alternative arguments, but with a number of important differences. 
These include a dependence on the de Sitter horizon, which may be expressed in terms of the cosmological constant as $l_{\rm dS} \sim 1/\sqrt{\Lambda}$. Applying the DE-UP to both charged and neutral particles, we obtain estimates of two limiting mass scales, expressed in terms of the fundamental constants $\left\{G,c,\hbar,\Lambda, e\right\}$. 
Evaluated numerically, the charged particle limit corresponds to the order of magnitude value of the electron mass ($m_e$), while the neutral particle limit is consistent with current experimental bounds on the mass of the electron neutrino ($m_{\nu_e}$). 
Possible cosmological consequences of the DE-UP are considered and we note that these lead naturally to a holographic relation between the bulk and the boundary of the Universe.
Low and high energy regimes in which dark energy effects may dominate canonical quantum behaviour 
are identified and the possibility of testing the model using near-future experiments is briefly discussed. 


\end{abstract}



\maketitle

\tableofcontents

\section{Introduction} \label{sect1}

The concept of superposition is the very essence of quantum theory. 
As the mathematical embodiment of wave-particle duality, it determines the state space structure of canonical non-relativistic quantum mechanics (QM) and its relativistic extension, quantum field theory (QFT). 
However, despite the unparalleled success of both QM and QFT in describing the micro-world, such duality does not manifest itself in our every day experience: the macro-world does not admit superpositions of states. 
This gives rise to the so-called measurement problem, whereby a classical `observer' (an experimenter or apparatus not subject to the quantum formalism) is required to reduce the quantum superposition via the act of `measurement'. 

With the measurement problem in mind, it is natural to consider the weakness of gravity, as compared to the three other known fundamental forces. 
Indeed, {\it classical} gravitational interactions may typically be ignored in the micro-world and only become relevant on macroscopic, even astrophysical or cosmological, scales \cite{Davies:1979mj}. 
Nonetheless, the exact nature of quantum gravitational interactions is unknown and their description remains the holy grail of theoretical physics research \cite{Kiefer:2007mf,Ellis:1999rz}. 
It is therefore natural to suppose that what is missing from canonical quantum theory is not an adequate description of the observer, vis-{\` a}-vis the observed, but gravity. 
Since the gravitational interaction is universal, affecting all forms of matter and energy, it may be hoped that gravity, or space-time itself, may play a fundamental role in the `spontaneous' decoherence of quantum systems. 

The idea that quantum gravitational effects may play an important role in the resolution of the measurement problem has a long history \cite{Karolyhazy:1966zz,KFL,Diosi:1988uy,Penrose:1996cv,Singh:2015sua}.
Originally published in 1966, K{\' a}rolyh{\' a}zy's model \cite{Karolyhazy:1966zz,KFL} was one of the first to consider the possibility of gravitationally-induced wave function collapse. 
The fundamental idea proposed in \cite{Karolyhazy:1966zz} is that quantum fluctuations of the metric give rise to an intrinsic and irremovable `haziness' in the space-time background, corresponding to a superposition of classical geometries. 
As a result, an initially pure state vector develops, over time, into a mixed state. 
Coherence is maintained only over a small region, known as a coherence cell, whose size depends on the space-time curvature induced by the body and, hence, on its mass. 
For micro-objects, the effect of curvature is small, giving rise to canonical quantum behaviour but, for macro-objects, the maximum size of a coherence cell lies within the classical radius of the body itself. 
Thus, the quantum nature of the macro-body remains hidden, as the wave function associated with its centre of mass (CoM) spontaneously decoheres on extremely small scales: the larger the body, the smaller the size of the cell.

From a theoretical perspective, a major advantage of the K{\' a}rolyh{\' a}zy model is that it contains no free parameters. 
It is therefore able to make clear predictions regarding gravitational modifications of the canonical quantum dynamics, utilising only the known constants $G$, $c$ and $\hbar$. 
Specifically, the existence of a minimum length uncertainty relation (MLUR), representing a modification of the canonical Heisenberg uncertainty principle (HUP), {\it necessarily} follows from the intrinsic haziness of space-time assumed in the K-model. 
The resulting uncertainty, inherent in the measurement of a space-time interval $s$, is 
\begin{eqnarray} \label{Karolyhazy_MLUR_Intro}
\Delta s \gtrsim (l_{\rm Pl}^2s)^{1/3} \, ,
\end{eqnarray}
where $l_{\rm Pl} = \sqrt{\hbar G/c^3}$ is the Planck length  \cite{Karolyhazy:1966zz,KFL}. 
For space-like intervals, this represents the minimum possible uncertainty in the position of a quantum mechanical particle, used to `probe' the distance $s$. 
When $\Delta s$ is identified with the Compton wavelength, $\lambda_{\rm C} = \hbar/(mc)$, $s$ may be identified with K{\' a}rolyh{\' a}zy's estimate of the width of a coherence cell for a fundamental particle, $a_{\rm c} \simeq  \lambda_{\rm C}^3/l_{\rm Pl}^2$.

Though motivated by an attempt to resolve the measurement problem, the MLUR (\ref{Karolyhazy_MLUR_Intro}) represents an important theoretical prediction in its own right. 
Since its inception, the literature on quantum gravity phenomenology has expanded significantly and many modifications of the HUP, known as generalised uncertainty principles (GUPs), have been proposed \cite{Tawfik:2015rva,Tawfik:2014zca}. 
These share the common feature of giving rise to a minimum resolvable resolvable length in nature, which is usually assumed to be of the order of the Planck length. 
Hence, the existence of some form of MLUR is now regarded as a generic feature of candidate quantum gravity models \cite{Hossenfelder:2012jw,Garay:1994en}. 

In this work, we will not concern ourselves with the measurement problem {\it per se}. 
Instead, we will focus on the second major prediction stemming from the introduction of a `hazy' space-time, i.e., that of a fundamental MLUR in nature. 
In particular, we will focus on a major advance in fundamental physics, which should have radical implications for {\it any} model quantum gravity, including MLURs, namely, the discovery of dark energy \cite{Reiss1998,Perlmutter1999}. 

Though the precise microphysical origin of dark energy remains unknown, and is an active area of research within the cosmology / astrophysics community, the current best-fit to {\it all} available cosmological data favours a `cosmological concordance' or $\Lambda$CDM model \cite{Ostriker:1995rn}, in which dark energy takes the form of a positive cosmological constant, $\Lambda > 0$. 
This accounts for approximately 69$\%$ of the total energy density of the Universe, whereas cold dark matter (CDM) accounts for around 26$\%$ and ordinary visible matter for around 5$\%$ \cite{Betoule:2014frx,PlanckCollaboration}.
For our purposes, it is important to note that, although dynamical dark energy models cannot be excluded on the basis of presently available data, {\it any} viable dark energy model must give rise to an effective cosmological constant at late times, comparable to the present epoch \cite{AmendolaTsujikawa(2010),Li:2012dt}. 
Furthermore, though $\Lambda$ may, ultimately, correspond to a form of matter in the usual sense (albeit of an exotic kind), its precise origin is unimportant for the derivation of dark energy-modified MLURs. 
What {\it is} important are its gravitational effects. 

Hence, $\Lambda$ may be interpreted as a {\it minimum space-time curvature}, or minimum gravitational field strength. 
This clearly has implications for any MLUR purporting to include quantum gravity effects, including K{\' a}rolyh{\' a}zy's model, which originally assumed quantum fluctuations of asymptotically flat space \cite{Karolyhazy:1966zz,KFL}. 
By contrast, we embed a K-type model in a realistic background geometry, incorporating the effects of dark energy. 
A key consequence of the existence of a positive cosmological constant is the existence of a fundamental horizon for {\it all} observers (including quantum mechanical `particles'), the de Sitter horizon, $l_{\rm dS} \sim 1/\sqrt{\Lambda}$. 
We argue that this necessarily implies a modification of the MLUR (\ref{Karolyhazy_MLUR_Intro}), including minimum curvature / finite-horizon effects. 
As with the original model presented in \cite{Karolyhazy:1966zz,KFL}, our model has the theoretical advantage of involving {\it no free parameters}. 
The main difference is that the MLUR obtained by considering a hazy space-time, {\`a} la K{\' a}rolyh{\' a}zy, in the presence of dark energy, necessarily involves $G$, $c$, $\hbar$ {\it and} $\Lambda$.

The structure of this paper is as follows. 
In Sec. \ref{sect2.1}, we consider classical perturbations of the cosmological Friedmann-Lem{\^a}itre-Robertson-Walker (FLRW) metric, induced by the presence of point particles. 
Although the FLRW metric is not valid on local scales, we note that its perturbed form, at the present epoch, is similar to the Schwarzschild-de Sitter metric. 
Thus, it predicts approximately the same gravitational potential (up to numerical factors of order unity) in the vicinity of a local compact object. 
This allows us to view the local field -- for example, around a microscopic particle located close to the surface of the Earth -- as a perturbation away from the cosmological background geometry. 
Throughout our analysis, $\Lambda$ is treated as a fundamental constant of nature which gives rise to a constant dark energy density, and minimum curvature, at {\it all} points in space. 

In Sec. \ref{sect2.2}, we show how the formula for the perturbed line element relates to K{\' a}rolyh{\' a}zy's scheme for resolving time-like intervals traversed by gravitating particles and show how this may be generalised to resolve space-like intervals -- in principle, even up to the horizon distance $\sim 1/\sqrt{\Lambda}$. 
Secs. \ref{sect2.3.1}-\ref{sect2.3.2} review the derivations of Eq. (\ref{Karolyhazy_MLUR_Intro}) given in \cite{Karolyhazy:1966zz,KFL} and \cite{Ng:1993jb,Ng:1994zk}, respectively, while Sec. \ref{sect2.4} outlines physical reasons for expecting dark energy-induced modifications of the standard formula. 

The main physical assumption behind the DE-UP is laid out in Sec. \ref{sect3.1} and the final result is derived in \ref{sect3.2}. 
The basic properties of the DE-UP, including applications to both neutral and electrically charged particles, are considered in Secs. \ref{sect3.3.1}-\ref{sect3.3.2}. 
Sec. \ref{sect4} considers possible cosmological implications and Sec. \ref{sect5} contains a summary of our conclusions together with a brief discussion of prospects for future work. 
Potential conceptual issues regarding the limits of applicability of the model, which arise at various points throughout the text, are discussed at greater length in the Appendix. 

\section{K{\' a}rolyh{\' a}zy's MLUR -- new perspectives} \label{sect2}

In \cite{Karolyhazy:1966zz,KFL}, K{\' a}rolyh{\' a}zy {\it et al} consider `resolving' a space-time interval $s$, traversed by a quantum mechanical particle of mass $m$, by projecting it into the lab frame using light signals emitted by the particle over the course of its path. 
They claim that {\it classically}, the observed interval $s'$ is related to the original (`true') interval $s$ via
\begin{eqnarray} \label{Karolyhazy_s'}
s' \simeq \left(1 - \frac{r_{\rm S}}{2r}\right)s \, , 
\end{eqnarray}
where $r_{\rm S}(m) = 2Gm/c^2$ is the Schwarzschild radius associated with the mass $m$. 
By explicitly taking into account the quantum nature of the particle traversing $s$, they then obtain an estimate of the minimum uncertainty in the measurement of $s$, denoted $\Delta s$ (\ref{Karolyhazy_MLUR_Intro}). 

In Sec. \ref{sect2.1}, we show that a formally similar result, in which the quantities $s$ and $s'$ in (\ref{Karolyhazy_s'}) have different physical meanings, may be obtained using gravitational perturbation theory. 
In this formulation, the quantities $s$ and $s'$ do not {\it a priori} represent `true' (CoM frame) and `measured' (lab frame) values of the length of a space-time interval but, instead, the lengths of an interval in an unperturbed background space and in the perturbed space induced by the presence of the particle, respectively. 
Nonetheless, the new formulation may be reconciled with K{\' a}rolyh{\' a}zy's picture, and the formal equivalence of the two pictures is shown explicitly in Sec \ref{sect2.2}. 

The original derivation of the MLUR (\ref{Karolyhazy_MLUR_Intro}) from Eq. (\ref{Karolyhazy_s'}), given in \cite{Karolyhazy:1966zz,KFL}, is considered in Sec. \ref{sect2.3.1} and K{\' a}rolyh{\' a}zy's measurement procedure is illustrated in Figs. 1-2.  
In Sec. \ref{sect2.3.2}, we review an alternative derivation of Eq. (\ref{Karolyhazy_MLUR_Intro}), originally given in \cite{Ng:1993jb,Ng:1994zk}, that does not rely on Eq. (\ref{Karolyhazy_s'}). 
Possible shortcomings of both approaches are discussed in \ref{sect2.4}, where it is argued that dark energy-induced modifications naturally resolve some outstanding problems. 

\subsection{Classical intervals in perturbed and unperturbed backgrounds: $s'$ and $s$} \label{sect2.1}

We now consider the classical perturbation induced by the presence of a point particle in a realistic space-time background, requiring the perturbed metric to satisfy the linearised Einstein equations. 
By `particle' we mean a spherically symmetric compact object that is point-like with respect to large -- in principle, up to cosmological -- length-scales. 

In the presence of dark energy, represented by a positive cosmological constant $\Lambda > 0$, the gravitational action is  
\begin{eqnarray} \label{EHA}
S = \frac{c^4}{16\pi G}\int (R-2\Lambda)\sqrt{-g}d^4x
\end{eqnarray}
and the field equations take the form
\begin{eqnarray} \label{EFE}
G_{\mu\nu} + \Lambda g_{\mu\nu} = \frac{8\pi G}{c^4}T_{\mu\nu} \, , 
\end{eqnarray}
where $g_{\mu\nu}$ denotes the space-time metric, $G_{\mu\nu} = R_{\mu\nu} - (1/2)R g_{\mu\nu}$ is the Einstein tensor, $R_{\mu\nu}$ is the Ricci tensor, $R=g^{\mu\nu}R_{\mu\nu}$ is the scalar curvature and $T_{\mu\nu}$ is the matter energy-momentum tensor. For a perfect fluid, $T_{\mu\nu}$ may be represented covariantly as
\begin{eqnarray} \label{EMT}
T_{\mu\nu} = (\rho c^2 + p)u_{\mu}u_{\nu} - pg_{\mu\nu} \, ,
\end{eqnarray}
where $\rho$ denotes the rest-mass density, $p$ is the isotropic pressure and $u^{\mu}$ is the $4$-velocity of an infitesimal fluid element. 

The FLRW metric describing a homogenous, isotropic, expanding Universe, may be written as
\begin{eqnarray} \label{FLRW_metric}
ds^2 = c^2d\tau^2 - a^2(\tau)d\Sigma^2 \, ,
\end{eqnarray}
where $\tau$ is the cosmic time and $a(\tau)$ is the cosmological scale factor which is normalized to one at the present epoch, $a(\tau_0) = a_0=1$. 
In spherical polar coordinates, $d\Sigma^2 = (1-kr^2)^{-1}dr^2 + d\Omega^2$ where $d\Omega^2 = r^2(d\theta^2 + \sin^2\theta d\phi^2)$ is the line-element for the unit $2$-sphere and $k$ is the Gaussian curvature, with dimensions $[L]^{-2}$. 
In appropriate units, $k \in \left\{-1,0,+1\right\}$ for negative, zero, and positive curvature, respectively. 
Substituting Eqs. (\ref{EMT}) and (\ref{FLRW_metric}) into (\ref{EFE}) yields the well-known Friedmann equations
\begin{eqnarray} \label{FriedmannEq-1}
\left(\frac{\dot{a}}{a}\right)^2 + \frac{kc^2}{a^2} - \frac{\Lambda c^2}{3} = \frac{8\pi G}{3}\rho \, ,
\end{eqnarray}
\begin{eqnarray} \label{FriedmannEq-2}
\frac{\ddot{a}}{a} = -\frac{4\pi G}{3}\left(\rho + \frac{3p}{c^2}\right) + \frac{\Lambda c^2}{3} \, ,
\end{eqnarray}
where a dot represents differentiation with respect to $\tau$ \cite{Islam:1992nt}. 
For future reference, we note that the Hubble parameter as is defined as
\begin{eqnarray}  \label{Hubble_param}
{\cal H} = \dot{a}/a \, ,
\end{eqnarray} 
and that its present day value is ${\cal H}_0 = 67.74 \pm 0.46 \ \rm kms^{-1}Mpc^{-1}$, or ${\cal H}_0 = 2.198 \times 10^{-18} \ \rm s^{-1}$ (ignoring error bars) in cgs units \cite{PlanckCollaboration}. 

In an arbitrary spatial coordinate system, Eq. (\ref{FLRW_metric}) may be written in the general form
\begin{eqnarray} \label{FLRW_metric-covariant}
ds^2 = c^2d\tau^2 - a^2(\tau)\gamma_{ij}dx^{i}dx^{j} \, , 
\end{eqnarray}
where $\gamma_{ij}$ is the spatial part of the metric, and an arbitrary metric perturbation may be written as 
\begin{eqnarray}  \label{arbitrary_pert}
g_{\mu\nu} \rightarrow g_{\mu\nu}' = g_{\mu\nu} + h_{\mu\nu} \, .
\end{eqnarray}
The gauge invariant tensor perturbations (`gravitons') satisfy the transverse-traceless conditions, $h^{i}{}_{i}=0$, $\nabla_{i}h^{i}{}_{j} = 0$, where $\nabla_{i}$ is the covariant derivative for the three-dimensional metric $\gamma_{ij}$. 
Let us now switch back to spherical polar coordinates and consider a spherically symmetric perturbation, induced by the `birth' of a particle of mass $m$, at some time $\tau' < \tau_0$. Our ansatz for the perturbative part of the energy-momentum tensor $T'^{\mu}{}_{\nu}$ then takes the form
\begin{eqnarray} \label{particle_ansatz}
T'^{\tau}{}_{\tau}(r,\tau)  \propto \frac{m\delta(r)}{2\pi a^2(\tau) r^2}\Theta(\tau - \tau') \, ,
\end{eqnarray}
where $\Theta$ is the Heaviside step function and all other components are zero. 
Strictly, Eq. (\ref{particle_ansatz}) models the birth of a particle, at $\tau = \tau'$, which remains at rest with respect to a comoving coordinate system at all later times. It also holds {\it approximately} for particles that are not subjected to extreme accelerations. 
In this case, dynamical tensor perturbations, which would otherwise lead to gravitational wave emission, may be neglected. 
\footnote{Though we here confine our attention to non-relativistic particles, which may be considered as approximately comoving with the background expansion of the universe, and, hence, do not generate gravitational waves, this assumption may break down in the very early universe, close to the end of the inflationary era, when densities and pressures were far higher than at the present epoch. 
Thus, it is interesting to note that, in MLUR scenarios, the dispersion relation for gravitational waves may be modified, leading to potentially measurable MLUR-induced effects on the background power spectrum, as shown 
in \cite{Cai:2007xr,Cai:2014hja}.}
In addition, we may set $B_{i} = h_{0i} = 0$ since, at linear order, vector perturbations are associated with vorticity in the cosmic fluid and do not arise in this scenario \cite{Christopherson:2011ra}. 

The full evolution of the scalar and tensor-type perturbations for the birth of a point-like mass may be determined by following a procedure analogous to that used in \cite{Lake:2011aa}, though such a detailed treatment is unnecessary for our current purposes.
Instead, we note that the covariant metric (\ref{FLRW_metric-covariant}) contains four extraneous degrees of freedom associated with coordinate invariance. 
In the Newtonian gauge, which holds {\it approximately} for situations in which $h_{0i} \simeq 0$ and where wave-like tensor perturbations can be neglected, this `gauge' freedom may be used to diagonalise the perturbed metric, giving 
\begin{eqnarray}  \label{FLRW_pert}
ds'^2 = \left(1+\frac{2\Psi}{c^2}\right) c^2 d\tau^2 - \left(1- \frac{2\Phi}{c^2}\right) a^2(\tau) \gamma_{ij} dx^i dx^j 
\end{eqnarray}
where $\Psi$ and $\Phi$ are Newtonian potentials obeying Poisson's equation \cite{Christopherson:2011ra}. 

In the absence of anisotropic stresses, $\Phi = \Psi$ \cite{Christopherson:2011ra}, and Poisson's equation for a mass distribution $\rho_{\rm m}$ immersed in a constant dark energy background is  
\begin{eqnarray}  \label{Poisson_Eq_modified}
\tilde{\nabla}^2 \Phi = 4\pi G\rho_{\rm m} - \frac{4\pi}{3} G \rho_{\Lambda} \, ,
\end{eqnarray}
where 
\begin{eqnarray} \label{Lambda_E_p}
\rho_{\Lambda} = -p_{\Lambda}/c^2 = \frac{\Lambda c^2}{8\pi G} 
\end{eqnarray}
is the dark energy density and $\tilde{\nabla}^2$ is the Laplacian, defined with respect to comoving coordinates. For spherically symmetric systems, this reduces to $\tilde{\nabla}^2 = (1/R^2)\partial_{R}(R^2\partial_{R})$, where 
\begin{eqnarray}  \label{comoving_coord}
R(\tau) = a(\tau)r \, .
\end{eqnarray}
The current experimental value of $\Lambda$, inferred from observations of high-redshift type 1A supernovae (SN1A), Large Scale Structure (LSS) data from the Sloan Digital Sky Survey (SDSS) and Cosmic Microwave Background (CMB) data from the Planck satellite, is $\Lambda = 1.114 \times 10^{-56} \rm cm^{-2}$ \cite{Betoule:2014frx,PlanckCollaboration}. 
This is equivalent to the vacuum energy density $\rho_{\Lambda} = 5.971 \times 10^{-30} {\rm g cm^{-3}}$. 

Now let us consider the case in which $\rho_{\rm m}$ is given by a $\delta$-function density profile corresponding to a classical point-like mass $m$, $\rho_{m}(\tau,r) \propto m\delta(r)/a^2(\tau)r^2$. 
This is equivalent to assuming the perturbed energy-momentum tensor (\ref{particle_ansatz}).
In this scenario, Eq. (\ref{Poisson_Eq_modified}) is simply Poisson's equation with two source terms, a regular point-mass ($m > 0$) and an `irregular' constant {\it negative} density, $-\rho_{\Lambda}$. 
(Recall that, when written on the right-hand side of the field equations, $\Lambda$ may be interpreted as a negative energy density belonging to the matter sector.) 
This is satisfied by the modified Newtonian potential
\begin{eqnarray}  \label{Modified_Newtonian_potential} 
\Phi(\tau,r) = -\frac{Gm}{ar} - \frac{\Lambda c^2}{6}a^2r^2 \, .
\end{eqnarray}
which gives rise to the gravitational field strength \cite{Hobson:2006se}
\begin{eqnarray}  \label{Modified_Newtonian_g}
\vec{g}(\tau,r) = -\vec{\tilde{\nabla}}\Phi(\tau,r) = \left(-\frac{Gm}{a^2r^2} + \frac{\Lambda c^2}{3}ar\right) \hat{\vec{r}} \, .
\end{eqnarray}

Thus, the cosmological constant corresponds to an effective gravitational repulsion whose strength increases linearly with the comoving distance $ar$ and we note that the force between two particles is attractive (repulsive) for $r \leq (\geq) r_{\rm grav}$, where 
\begin{eqnarray}  \label{Turn_Around_Radius_l_W}
r_{\rm grav}(\tau) = 2^{-1/3}a^{-1}(\tau)(l_{\rm dS}^2r_{\rm S})^{1/3} \, ,
\end{eqnarray}
and $l_{\rm dS} = \sqrt{3/\Lambda} = 1.641 \times 10^{28}$ cm. $l_{\rm dS}$ is the asymptotic de Sitter horizon and is of the same order of magnitude as the present day radius of the Universe $r_{\rm U} \simeq 1.306 \times 10^{28}$ cm (13.8 billion light years). 
In the Newtonian picture, $r_{\rm grav}$ marks the separation distance beyond which the effective gravitational force between two spherically symmetric bodies becomes repulsive, i.e., beyond which the repulsive effect of dark energy overcomes the canonical gravitational attraction. 

Including contributions to $\rho_{m}$ from the background baryonic and dark matter densities -- that is, embedding the perturbation in a full FLRW background geometry -- similar arguments yield 
\begin{eqnarray}  \label{Modified_Newtonian_potential-BETTER}
\Phi(\tau,r) = -\frac{Gm}{ar} - \frac{{\cal H}^2}{2}a^2r^2 \, ,
\end{eqnarray}
where ${\cal H(\tau)}$ is a solution to Eqs. (\ref{FriedmannEq-1})-(\ref{FriedmannEq-2}), so that 
\begin{eqnarray}  \label{Turn_Around_Radius}
r_{\rm grav}(\tau) = 2^{-1/3}a^{-1}(\tau)(r_{\rm S}c^2/{\cal H}^2(\tau))^{1/3} \, . 
\end{eqnarray}
This is known as the gravitational turn-around radius, and may also be derived rigorously in a fully general relativistic context \cite{Bhattacharya:2016vur}. 
For $\tau \rightarrow \infty$, the matter density is diluted to such an extent that ${\cal H}^2 \rightarrow c^2/l_{\rm dS}^2$ and the space-time becomes asymptotically de Sitter, yielding Eq. (\ref{Turn_Around_Radius_l_W}). 

We note that the inclusion of a $\Lambda$ / $\mathcal{H}$-dependent term inside the Newtonian potential does not violate the cosmological principle, since $\Phi$ describes only local perturbations. 
Indeed, our perturbative analysis is valid only on scales $r \lesssim r_{\rm grav}$ ($\Phi(r) \lesssim 0$), beyond which the gravitational influence of the point-mass may be ignored.   

In more complex local environments, we may expect $\mathcal{H}(\tau)$ appearing in Eq. (\ref{Modified_Newtonian_potential-BETTER}) to be replaced by a {\it local} Hubble parameter, $\mathcal{H}(\tau,\vec{r})$, which is not a solution of Eqs. (\ref{FriedmannEq-1})-(\ref{FriedmannEq-2}). 
However, if $\Lambda$ is a genuine constant of nature, giving rise to a constant dark energy density at {\it all} points in space (as assumed in this analysis), there still exists a correction term to the canonical Newtonian potential, determined by the local Hubble parameter, $\mathcal{H}/c \geq 1/l_{\rm dS}$. 

Hence, in general, the infinitesimal line-elements of the perturbed and unperturbed metrics, $ds'$ and $ds$, are related via
\begin{eqnarray} \label{FLRW_pert_static-1}
ds' &=& \sqrt{1 - \frac{r_{\rm S}}{ar} - \frac{{\cal H}^2 a^2r^2}{c^2}}ds 
\nonumber\\
&\simeq&  \left(1 - \frac{r_{\rm S}}{2ar} - \frac{{\cal H}^2a^2r^2}{2c^2}\right)ds \, ,
\end{eqnarray}
where minimum value of the Hubble term is set by the dark energy scale. 
Here, $ds$ denotes the line-element for the flat, unperturbed, space-time. Next, we rewrite the unperturbed line-element Eq. (\ref{FLRW_metric-covariant}) as $ds^2 = c^2d\tau^2[1 - a^{2}(\tau)\gamma_{ij}(dx^{i}/d\tau)(dx^{j}/d\tau)]$. 
Restricting ourselves to time-like intervals within the present day horizon then gives 
\begin{eqnarray} \label{ds=cdtau}
ds \simeq cd\tau \, .  
\end{eqnarray}
This assumption is valid if $x^i(\tau)$ represents the embedding of a non-relativistic particle. Substituting Eq. (\ref{ds=cdtau}) into Eq. (\ref{FLRW_pert_static-1}) then allows us to obtain a {\it lower bound} on $s'(\tau,r)$, such that
\begin{eqnarray} \label{FLRW_pert_static-2}
s'(\tau,r) \gtrsim  \left(1 - \frac{r_{\rm S}}{2a r} - \frac{{\cal H}^2a^2r^2}{2c^2}\right)s(\tau) \, ,
\end{eqnarray}
where $s(\tau) = c\tau$. 
Most importantly, for $\tau \simeq \tau_0$, which implies $a \simeq a_0 =1$ and ${\cal H}^2 \simeq {\cal H}_0^2$, the final term is subdominant for $r \lesssim r_{\rm grav}(\tau_0) = (c^4/16\pi G)^{-1/3}(3r_{\rm S}/\rho_{\rm crit}c^2)^{1/3}$, so that Eq. (\ref{FLRW_pert_static-2}) reduces to Eq. (\ref{Karolyhazy_s'}) in this regime. 

Hence, we have recovered the starting point for K{\' a}rolyh{\' a}zy's analysis, Eq. (\ref{Karolyhazy_s'}).
However, here, $s'$ and $s$ denote perturbed and unperturbed line-elements, not lab frame and CoM frame intervals, respectively. 
We now demonstrate the equivalence of these two pictures. 

\subsection{Classical equivalence of K{\' a}rolyh{\' a}zy's measurement scheme and the perturbative result: reinterpreting $s'$ and $s$} \label{sect2.2}

\subsubsection{Asymptotically flat space ($\Lambda = 0$)} \label{sect2.2.1}

Neglecting the sub-dominant Hubble flow term in Eq. (\ref{FLRW_pert_static-2}), the intervals $s'$ and $s$ in may also be interpreted in terms of the following experimental procedure. 
Suppose we place a `detector' at a coordinate distance $r > 0$ from a specified origin.  
(We assume throughout that our detector represents an idealized observer whose gravitational field may be considered negligible, even compared to that of the perturbing particle, which is located at $r=0$. 
Though unrealistic, this is a valid assumption in our idealized gedanken experiment.) 
If the mass of the particle may be neglected, a photon travelling from $r=0$ to the detector at $r > 0$ traverses a space-like interval $s(r) = r \equiv s(\tau) = c\tau$, where $\tau$ is the flight-time.
(Of course, the photon also traverses a time-like interval $c\tau$, so that its total trajectory is null. However, for our proposed measurement procedure, only the space-like interval is relevant.) 
By contrast, if the mass of the particle is non-negligible at $r$, a photon emitted at $r=0$ and absorbed by the detector after a time $\tau$ traverses a space-like interval $s'(r,\tau)$ (\ref{FLRW_pert_static-2}) as seen by the observer at $r$. 
Thus, the simple relationship between the coordinate distance and the space-like interval is destroyed by the gravitational field of the particle. 

Furthermore, $\tau$ need not correspond to the flight time of single photon. 
Instead, we may consider spitting the measurement of the interval $s'(r,\tau)$ into two (or more) parts. For simplicity, however, we consider only a two part measurement process. 
In the first part, a photon travels from the perturbing particle at $r=0$ to the detector at $r >0$. In the second, an additional photon travels from a (generally different) point, to $r$. 
If the total flight time of {\it both} photons is $\tau$, the space-like interval that would have been traversed {\it if} the particle had not been present is $s(\tau) = c\tau$, but the interval traversed in the perturbed space is $s'(r,\tau)$. 
In other words, because the flight time of the photons and the position of the observer relative to the mass $m$ are {\it independent}, so are $r$ and $s(\tau)$. 
Since $r$ can label any point in space, regardless of the value of $\tau$, 
it follows that the measured interval $s'(r,\tau)$ depends on where we place our detector in relation to the particle. 

This fact also enables us to reinterpret Eq. (\ref{FLRW_pert_static-2}) in terms of an experimental procedure to resolve time-like intervals traversed by massive, self-gravitating particles, {\`a} la K{\' a}rolyh{\' a}zy. 
During the photon flight time $\tau$, the CoM of a {\it classical} non-relativistic particle also traverses a time-like interval approximately equal to $s(\tau) = c\tau$. Hence, $s'(r,\tau)$ and $s(\tau)$ may be interpreted as the `observed' (lab frame) and `true' (CoM frame) values of the time-like 
interval traversed by a massive particle, as claimed in \cite{KFL}. 
This procedure is illustrated in Figs. 1-2.

\subsubsection{Finite-horizon effects ($\Lambda > 0$)} \label{sect2.2.2}

Let us now consider such a two-part measurement in more detail, including the effects of a finite horizon distance. 
We again begin by assuming that the gravitational effect of the particle mass can be neglected, so that $s' \simeq s$ in our notation. 
This scenario is represented by the flat blue line in Fig. 1. 
Consider measuring a space-like distance by means of a photon, emitted from the particle at $r=0$ and absorbed by a detector in the lab frame at some distance $r = ct$, where $t$ is the proper time measured by particle's CoM. 
In general, this need {\it not} be identified with the cosmic time $\tau$, so that we are free to consider $t \lesssim \tau$. 
If the particle's recoil velocity is non-relativistic, it may be considered negligible at the {\it classical} level, so that $dt \simeq d\tau$. 
Thus, if $t$ is small compared to the cosmic time ($t \ll \tau$), we may set $a \simeq a_0 = 1$. 
In this case, it is clear that the {\it time-like} interval traversed by the particle in time $t$ is identical to the {\it space-like} interval measured by the experimental apparatus (i.e the particle-photon-detector system). 
In K{\' a}rolyh{\' a}zy's notation, we have $s'(t) = s(t) = ct \equiv s'(r) = s(r) = r$, where $s'$ and $s$ denote the world-lines traversed by the particle and measured in the lab frame, respectively.

Now let us consider the more general case, in which the space-time curvature induced by the presence of the particle cannot be ignored. 
This scenario is represented by the curved red line in Fig. 1. 
In this case, if the photon travels from the particle at $r=0$ to the detector at $r > 0$ in time $t$, this corresponds to the measurement of a {\it space-like} interval $s'(t,r) \simeq (1-r_{\rm S}/2r)ct$. 
As stated above, once the gravitational field of the particle is taken into account, the simple relation between the coordinate distance $r$, traversed by the photon, and the space-like interval this corresponds to breaks down ($s'(r) \neq r$). Likewise, the simple relationship between the coordinate distance and the time elapsed no longer holds ($r(t) \neq ct$). 

The {\it time-like} interval traversed by the particle is still $s(t) = ct$, so that $s'(t,r) \simeq (1-r_{\rm S}/2r)s(t)$, as in Eq. (\ref{Karolyhazy_s'}). 
However, $s(t) = ct$ also represents the {\it space-like} interval that {\it would} have been measured, had the particle's mass not perturbed the background. Hence, K{\' a}rolyh{\' a}zy's interpretation of the symbols $s$ and $s'$, as representing the `true' (CoM) frame and measured (lab frame) values of the space-time interval traversed by the particle, is {\it equivalent} to ours, in which they represent intervals in the non-perturbed and perturbed backgrounds, respectively.   

\begin{figure}[h]
     \label{fig1}\centering
      \includegraphics[width=8.7cm]{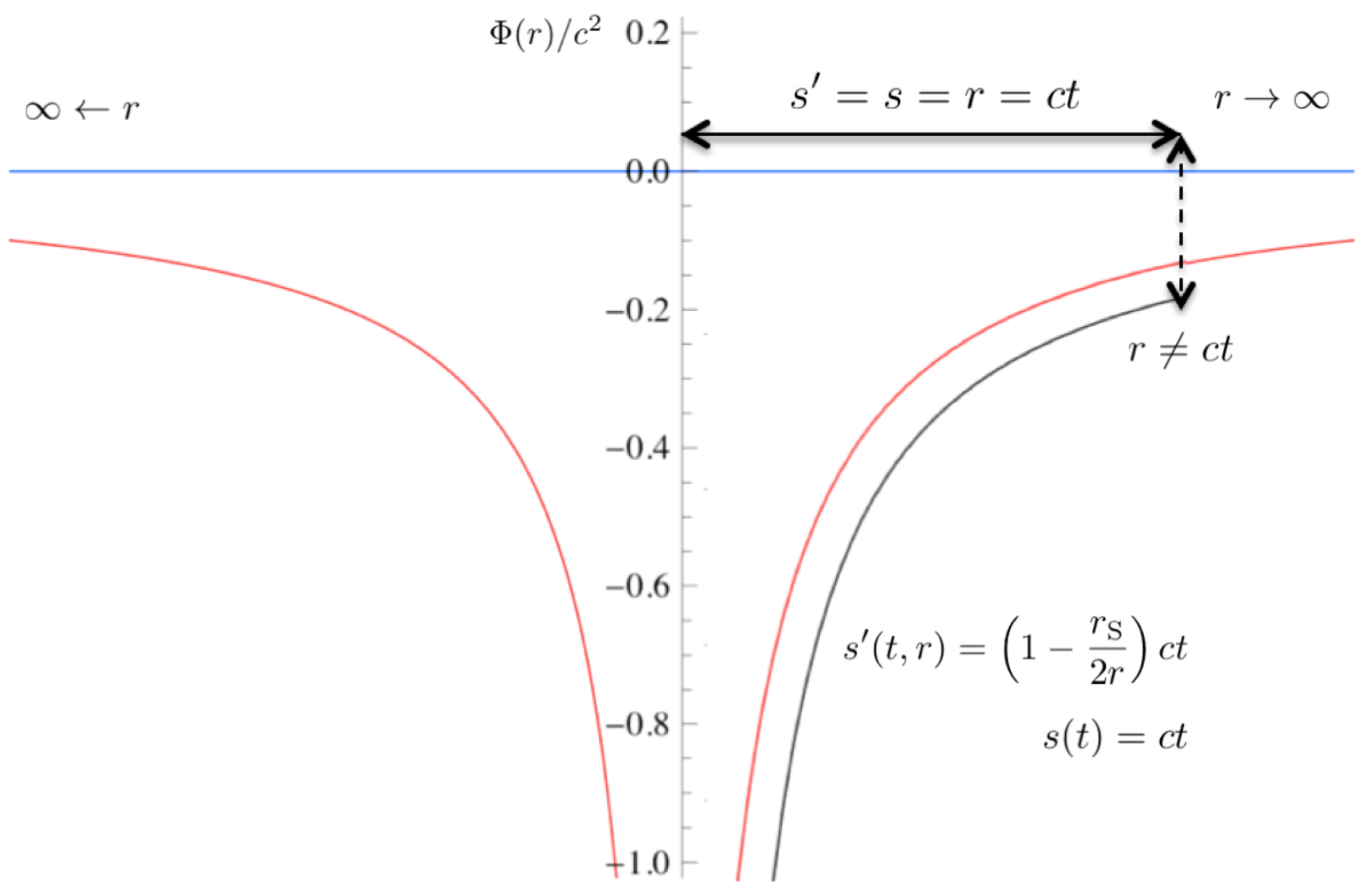}
     \caption{
     If the gravitational field of the particle is considered negligible, space-time is approximately flat (blue line). 
     In this case, a photon emitted from the particle at $r=0$ travels to a detector at the point $r(t)=ct$ in time $t$. 
     This completes a measurement of the space-like interval $s(t) = ct$. During this time (ignoring recoil), the particle traverses a time-like interval $s'(t) = ct$, so that $s'(t) = s(t) = r(t) = ct$. 
     Taking the particle's gravity into account (red line), if the photon travels from $r=0$ to $r>0$ in time $t$, this corresponds to a measurement of the space-like interval $s'(t,r) \simeq (1-r_{\rm S}/2r)ct$, $(r \neq ct)$. 
     The time-like interval traversed by the particle is still $s(t)=ct$, so that $s'(t,r) \simeq (1-r_{\rm S}/2r)s(t)$. 
     This formula relates the perturbed line element $s'(t,r)$ to the unperturbed line element $s(t)$ or, equivalently, the space-like interval measured at $r$ to the `true' time-like interval traversed by the particle. 
     Hence, the relation between $s'$ and $s$ obtained from K{\' a}rolyh{\' a}zy's measurement procedure, using {\it classical} particles, is equivalent to the perturbative result. 
     This procedure may be generalised to measure larger distances via multi-photon absorption at $r$.}
\end{figure}

\begin{figure}[h]
     \label{fig2}\centering
      \includegraphics[width=8.7cm]{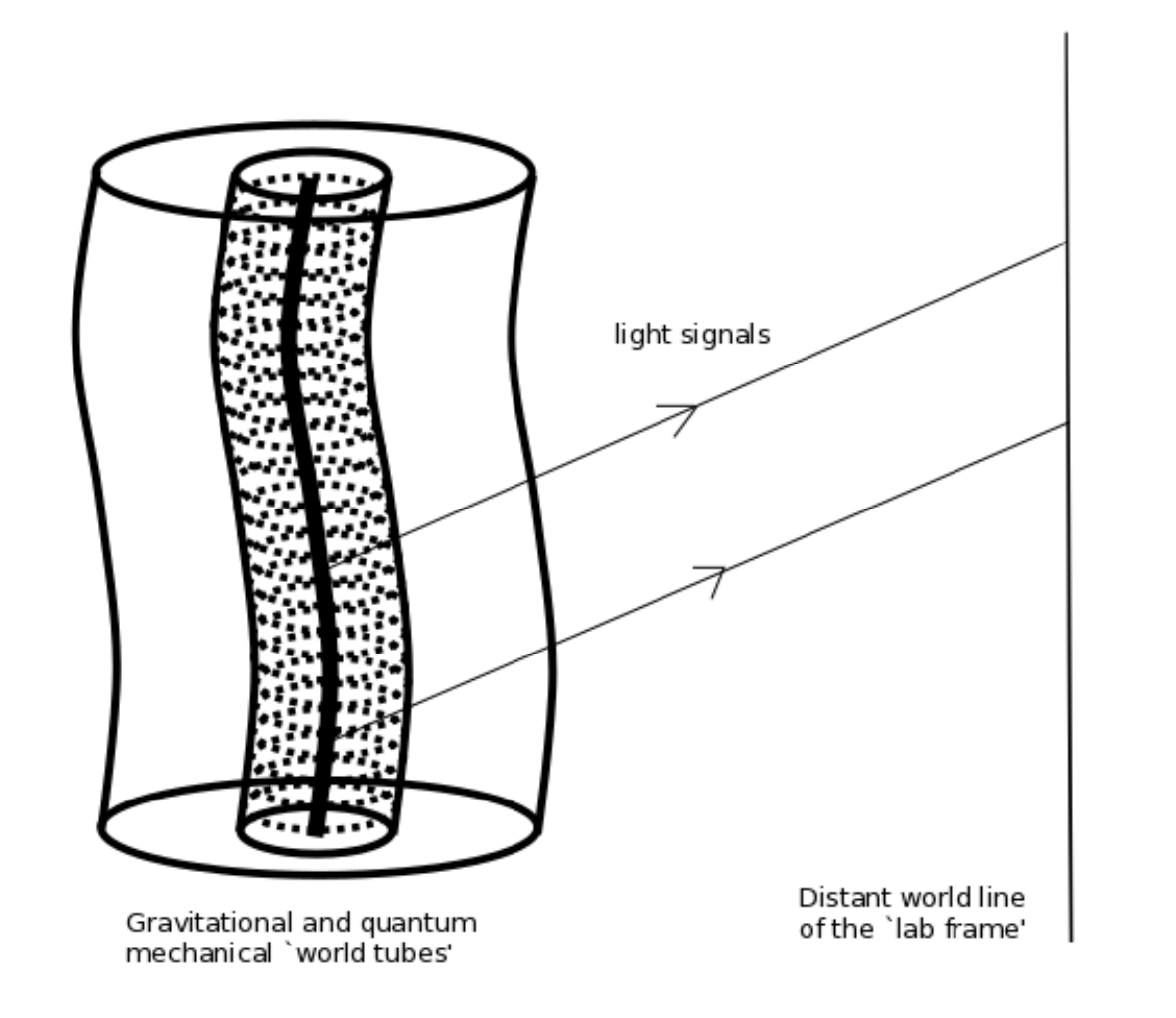}
     \caption{
     Measurement of the time-like interval traversed by a massive particle located at $r \simeq 0$, by projecting light-like signals emitted over the course of its path onto a `detector' at $r>0$. 
     The outer tube surrounding the centre of mass (CoM) represents the region $r < r_{\rm grav}$, in which the particle's gravitational field may be considered non-negligible compared to the background curvature. 
     Placing the detector within $r_{\rm grav}$ leads to significant differences between the measured (lab frame) and `true' (CoM frame) values, even in the classical regime (\ref{FLRW_pert_static-2}). 
     The inner tube represents the fuzziness of the particle's CoM due to the minimum width of the canonical quantum wave packet, i.e., the Compton radius. 
     Generally, the tubes defined by the gravitational and quantum mechanical radii have different thicknesses, but coincide for the minimum-mass particle predicted by the DE-UP. (See Sec. \ref{sect3.3}.)}
\end{figure}  

We now show, explicitly, that Eq. (\ref{Karolyhazy_s'}) holds even more generally. 
Suppose that, rather than measuring the space-like interval between the particle and the detector -- which corresponds to the coordinate distance $r$, even if the two are not equivalent -- we instead choose to measure a much larger interval. 
For example, let us imagine that the particle is surrounded by a horizon, at a fixed distance $s'$ from its CoM. 
Furthermore, let us imagine that, if the gravitational field of the particle were absent, the horizon would be located at a fixed distance $s = l_{*}$ rather than $s'$. 

Our experimental procedure is then as follows. 
A photon is emitted from the particle at $r=0$ and absorbed by the detector in the lab frame (as before) after a time $t_1$. 
This completes a measurement of the space-like interval $s_1' \simeq (1-r_{\rm S}/2r)s_1$, where $s_1 = c t_1$. Simultaneously, or near simultaneously (i.e., within a time interval $\sim r/c$), a photon emitted from a point on the horizon at $t_2=t_1-t_{*}$, where $t_{*} = l_{*}/c$, also arrives at $r$ and is absorbed by the detector. 
This completes a measurement of the space-like interval $s_2' \simeq (1-r_{\rm S}/2r)s_2$, where $s_2 = -ct_2 > 0$. This result follows directly from the independence of the space-time coordinates $r$ and $t$ where, in our experimental procedure, $t$ is identified with the flight time of a photon and $r$ is identified with the position of the detector. 
Together, these interactions complete the measurement of a space-like interval given by
\begin{eqnarray} \label{}
s'(r) = s_1' + s_2' \simeq \left(1 - \frac{r_{\rm S}}{2r}\right)l_{*} \, .
\end{eqnarray}
The time-like interval traversed by the particle during the flight time of {\it both} photons is $s = ct_{*} = l_{*}$, so that this procedure is equivalent to projecting the entire world-line of the particle, traced out over $t_{*}$, onto the detector at $r$.  

Modifying this argument to include the effects of universal expansion, dark energy ($\Lambda > 0$) and the background matter density on the Newtonian potential induced by the perturbation, gives 
\begin{eqnarray} \label{Horizon_measure-1}
s'(\tau, r) \gtrsim \left(1 - \frac{r_{\rm S}}{2a(\tau) r} - \frac{{\cal H}^2(\tau)a^2(\tau)r^2}{2c^2}\right) \frac{\mathcal{H}(\tau)}{c} \, , 
\end{eqnarray}
which is simply Eq. (\ref{FLRW_pert_static-2}) with $s(\tau)$ set equal to the Hubble horizon. 

Hence, the measured value of the space-like distance between the particle and the horizon depends on where we place our detector in relation to each. 
This is a simple consequence of the fact that the perturbation breaks the global symmetry (i.e. homogeneity or, equivalently, isotropy about every point) of the FLRW background. 
If $r$ is very small, the detector sits within a (relatively) deep potential well, in which the difference between the curvature of the perturbed and the unperturbed backgrounds is large. 
From K{\' a}rolyh{\' a}zy's viewpoint, the time-like interval traversed by the particle, over the time taken for a photon to reach the horizon, is projected onto a detector in the lab frame at $r$. 
If $r \lesssim r_{\rm grav}(\tau) \ll r_{*}$, where $r_{*}$ is the coordinate distance corresponding to the position of the horizon, the distortion induced by the gravitational field of the particle renders the measured value significantly different from the true (CoM frame) value. 

Implicitly, this argument assumes that the particle formed in the very early Universe ($\tau \simeq 0$). 
However, even if this is {\it not} the case, ${\cal H}(\tau)/c$ still marks the furthest point in causal contact with the particle at the cosmic epoch $\tau$. 
As such, it {\it still} represents the largest distance that can be measured by means of the particle-photon-detector system, at time $\tau$. 
Hence, Eq. (\ref{Horizon_measure-1}) remains physically meaningful in relation to the gedanken experiment described above, in which the detector at $r$ receives signals from both the particle at $r=0$ and its horizon at ${\cal H}(\tau)/c$. 

Finally, we note that similar considerations hold, even for electrically neutral particles, whose non-gravitational interactions are mediated by massive short-range bosons. 
Though the only long-range force affecting a neutral particle is gravity, this is sufficient to maintain causal contact with its horizon, which corresponds to the exchange of real {\it or virtual} particles, as described above. 
Indeed, we may screen charged particles from electromagnetic interactions using a Faraday cage, but we can never screen {\it all} signal exchange between a particle and its horizon. 
Such exchanges constitute de facto measurements of the horizon distance $\mathcal{H}(\tau)/c$, via the measurement procedure outlined above.  

\subsection{Probing space-time intervals with quantum particles} \label{sect2.3}

The above argument demonstrates the {\it classical} equivalence of K{\' a}rolyh{\' a}zy's measurement scheme and the perturbative result, Eq. (\ref{FLRW_pert_static-2}). 
In canonical QM, the picture of the classical point-particle is replaced by the wave function $\psi$, representing a superposition of position or, equivalently, momentum states of the particle's CoM. 
Thus, it is not difficult to imagine that, in the quantum regime, the classical region over which the particle's self-gravity {\it cannot} be neglected gives rise to an irreducible haziness of the underlying space-time metric, induced by the presence of the wave function. 
This is equivalent to an irreducible `smearing' out of the particle mass or, equivalently, of the CoM associated with $\psi$. 
This observation, which formed the basis of K{\' a}rolyh{\' a}zy's predictions \cite{Karolyhazy:1966zz,KFL}, will also form the basis of our own analysis, though we will depart from his original prescription in a number of ways. 
In particular, we will attempt to incorporate the effects of a space-filling dark energy, which exists in the form of a cosmological constant $\Lambda >0$, with effective energy density and pressure given by Eq. (\ref{Lambda_E_p}). 

On the one hand, we may restrict our attention to a very small region in the vicinity of the CoM, over which the particle's (extremely small) self-gravity may be considered non-negligible compared to the background level, {\it whatever} this may be. 
Classically, such a region is well defined for {\it any} perturbed metric and traces out a `world-tube' of width $r_{\rm grav}$ (\ref{Turn_Around_Radius}) surrounding the CoM world-line \cite{Karolyhazy:1966zz,KFL}. 
Projecting the world-line onto a `detector' within this tube gives rise to significant deviations in the measured value of the interval, as compared to its `true' value, due to the space-time curvature induced by the particle. 
On the other hand, since the dark energy density gives rise to both $r_{\rm grav}$ {\it and} the finite size of the horizon $\sim l_{\rm dS}$, our model considers the influence of the global horizon on {\it local} systems.

Clearly, once the `fuzziness' of the CoM due to canonical quantum mechanics is taken into account things become even more complicated, as a second radius -- the Compton radius -- may be associated with the particle. Nonetheless, 
as we will show explicitly in Sec. \ref{sect3}, the counter-intuitive results implied by the considerations above remain the same: once the particle's self-gravity is taken into account, physical measurements of space-time intervals -- for example, the space-like position of a particle, relative to a predefined origin -- yield {\it more accurate} results if the measurements are made from {\it further away}. 
Below a certain optimum length-scale, attempting to probe the position of the particle's CoM with greater accuracy becomes self-defeating. The resulting `gravitational uncertainty' caused by the fuzziness of the space-time close to the particle's CoM outweighs the gain in localising the canonical quantum wave packet. 
By contrast, far away from the CoM, metric fluctuations reduce to the background level (assumed to be of the order of the Planck length) and canonical quantum behaviour is recovered. 
The measurement scheme considered above is shown, for particles with both classical gravitational (turn-around) and quantum mechanical (Compton) radii, in Fig. 2. 

Next, we review K{\' a}rolyh{\' a}zy's original derivation of Eq. (\ref{Karolyhazy_MLUR_Intro}) (Sec. \ref{sect2.3.1}) together with an alternative argument that leads to the same result (Sec. \ref{sect2.3.2}). 
We then outline possible shortcomings of these models and give physical reasons for their modification in the presence of dark energy (Sec. \ref{sect2.4}). 

\subsubsection{K{\' a}rolyh{\' a}zy's MLUR (1968)} \label{sect2.3.1}

To highlight both the similarities and the differences between the arguments presented in \cite{Karolyhazy:1966zz,KFL} and those presented in the present work, we briefly review the original derivation of K{\' a}rolyh{\' a}zy's MLUR. Special emphasis is placed on the physical assumptions that underly the model and on the chain of reasoning that gives rise to the final result. For clarity, where new or supplementary assumptions are introduced for the first time, they are explicitly stated.

Beginning with Eq. (\ref{Karolyhazy_s'}), K{\' a}rolyh{\' a}zy effectively {\it defines} the uncertainty in $s'$ in terms of an {\it assumed} uncertainty in $m$, via
\begin{eqnarray} \label{ds'_K}
\Delta s' \simeq \beta \frac{G \Delta m}{c^2 r}s \, .
\end{eqnarray}
where $\beta$ is a positive numerical constant of order unity. In fact, following Eq. (\ref{Karolyhazy_s'}), $\beta$ is set exactly equal to one in K{\' a}rolyh{\' a}zy's original derivation \cite{Karolyhazy:1966zz,KFL}. We explicitly include it, from here on, for the sake of comparison with the results of Ng and van Dam \cite{Ng:1993jb,Ng:1994zk}. 

While this idea is reasonable from a gravitational perspective -- where one may expect statistical fluctuations in space-time configurations to be {\it equivalent} to fluctuations in the mass that `sources' the gravitational field (or at least correlated with them) -- it is problematic from the quantum point of view, since `uncertainty' refers to the statistical spread of measurement outcomes, where the physical quantity in question is represented by a Hermitian operator. 
However, in both canonical QM and QFT, mass is a parameter, {\it not} an operator. 

In \cite{Karolyhazy:1966zz,KFL}, K{\' a}rolyh{\' a}zy obtains the expression for $\Delta m$ from the `canonical' uncertainty relation $\Delta E \Delta t \gtrsim \hbar$, though this too is potentially problematic, as $t$ is not an operator in the canonical non-relativistic theory. 
Defining the uncertainty in the rest-energy of the particle as 
\begin{eqnarray} \label{dm_K-1}
\Delta E_{\rm rest} = \Delta m c^2 \, 
\end{eqnarray}
and using $s \simeq ct$ to infer $\Delta s \simeq c\Delta t$, yields 
\begin{eqnarray} \label{dm_K-2}
\Delta m \simeq \hbar/(c \Delta s) \, .
\end{eqnarray}
Substituting (\ref{dm_K-2}) into (\ref{ds'_K}), then assuming that the self-gravity associated with the particle's wave function is non-negligible only over the interval $0 \leq r \lesssim \Delta s$ [i.e., replacing $r \rightarrow \Delta r \simeq \Delta s$ in (\ref{ds'_K})] and noting that the minimal value of $\Delta s'$ is $(\Delta s')_{\rm min} \simeq \Delta s$, then yields 
\begin{eqnarray} \label{Karolyhazy_MLUR}
\Delta s \geq (\Delta s)_{\rm min} \simeq (\beta l_{\rm Pl}^2s)^{1/3} \, .
\end{eqnarray}

\subsubsection{Ng and van Dam's derivation (1994)} \label{sect2.3.2}

An alternative derivation of Eq. (\ref{Karolyhazy_MLUR}) is based on a gravitational extension of the MLUR obtained in canonical QM, and was originally proposed by Ng and van Dam \cite{Ng:1993jb,Ng:1994zk}. 
That an MLUR exists, even in the canonical non-gravitational theory, can be seen by considering the dependence of the positional uncertainty $\Delta x$ on the time interval $t$ over which measurements are made \cite{Calmet:2004mp,Calmet:2005mh}. (Note that we again distinguish between this and the cosmic time $\tau$.)

In the absence of an external potential ($V=0$), the time evolution of the position operator $\hat{x}(t)$ is given by 
\begin{eqnarray}  \label{Heis-1}
\frac{d\hat{x}(t)}{dt} = \frac{i}{\hbar}[\hat{H}(t),\hat{x}(t)] = \frac{\hat{p}(t)}{m} \, ,
\end{eqnarray}
which may be solved directly, yielding
\begin{eqnarray}  \label{Heis-2}
\hat{x}(t) = \hat{x}(0) + \hat{p}(0)\frac{t}{m} \, .
\end{eqnarray}
The spectra of any two Hermitian operators, $\hat{A}$ and $\hat{B}$, obey the general uncertainty relation \cite{Rae00,Ish95}
\begin{eqnarray}  \label{AB}
\Delta A \Delta B \geq \frac{1}{2}|\langle[\hat{A},\hat{B}]\rangle| \, ,
\end{eqnarray}
so that setting $\hat{A} = \hat{x}(0)$, $\hat{B} = \hat{x}(t)$ gives
\begin{eqnarray}
[\hat{x}(0),\hat{x}(t)] = i\hbar \frac{t}{m} \, ,
\end{eqnarray}
and
\begin{eqnarray}
\Delta x(0)\Delta x(t)  \geq \frac{\hbar t}{2m} \, .
\end{eqnarray}
Next, we define the uncertainty over {\it all} measurements, made at both $t=0$ and $t>0$, as the geometric mean of the canonical  uncertainties at both times, i.e.
\begin{eqnarray} \label{Delta_x_canon.}
\Delta x_{\rm canon.}(t) = \sqrt{\Delta x(0)\Delta x(t)} \, .
\end{eqnarray}
Using the definition of $\Delta x_{\rm canon.}(t)$, Eq. (\ref{Delta_x_canon.}), together with $t = r/c$, we obtain 
\begin{eqnarray} \label{Wiger-Salecker-1}
\Delta x_{\rm canon.}(t) \gtrsim \sqrt{\frac{\hbar t}{2m}} \equiv  \Delta x_{\rm canon.}(r) \gtrsim \frac{1}{\sqrt{2}}\sqrt{\lambda_{\rm C}r} \, ,
\end{eqnarray}

Historically, this result was first obtained by Salecker and Wigner using a gedanken experiment in which a quantum particle is used to measure a distance $r$ by means of the emission and reabsorption of a photon \cite{Salecker:1957be}. 
In this description $\Delta x_{\rm canon.}(r)$, given by Eq. (\ref{Wiger-Salecker-1}), represents the minimum possible {\it canonical} quantum uncertainty in the measurement of $r$. 

The argument presented in \cite{Salecker:1957be} proceeds as follows. Suppose we attempt to measure $r$ using a `clock' consisting of a classical mirror and a quantum mechanical device (e.g. a charged particle such as an electron), initially located at $r=0$, that both emits and absorbs photons. 
A photon is emitted at $t=0$ and reflected by the mirror, which is placed at some unknown distance $r>0$. The photon is then reabsorbed by the particle after a time $t = 2r/c$ ({\it not} $t = r/c$).

Assuming that the velocity of the particle remains well below the speed of light, it may be modelled non-relativistically. 
By the HUP the uncertainty in its velocity at any time $t \geq 0$ obeys the inequality
\begin{eqnarray}  \label{dv}
\Delta v(t) \gtrsim \frac{\hbar}{2m \Delta x(t)} \, ,
\end{eqnarray}
where $\Delta x(t)$ is the positional uncertainty obtained by evolving the initial wave function $\psi(x,0)$ via the Schr{\" o}dinger equation (i.e. neglecting recoil). 
However, if the initial positional uncertainty is $\Delta x(0)$, then, in the time required for the photon to travel to the mirror and back, $t$, the particle acquires an additional positional uncertainty 
\begin{eqnarray}  \label{dv}
\Delta x_{\rm recoil}(t) = \int_{0}^{t}\Delta v(t') dt' \gtrsim \Delta v(t) t \, .
\end{eqnarray}
The \emph{total} canonical positional uncertainty is now defined as 
\begin{eqnarray}  \label{recoil}
\Delta x_{\rm canon.}(t) = \Delta x(t) + \Delta x_{\rm recoil}(t) \, , 
\end{eqnarray}
and obeys the inequality
\begin{eqnarray}  \label{dx_tot-1}
&&\Delta x_{\rm canon.}(t) 
\gtrsim \Delta x(t) + \frac{\hbar t}{2m \Delta x(t)} 
\nonumber\\
&\equiv& \Delta x_{\rm canon.}(t) \gtrsim \frac{\hbar}{2 m\Delta v(t)} + \Delta v(t)t \, .
\end{eqnarray}
Minimizing this expression with respect to $\Delta x(t)$, or equivalently $\Delta v(t)$, and using the fact that $\Delta v_{\rm max} \simeq  \hbar/(2m \Delta x_{\rm min})$, gives
\begin{eqnarray}  \label{Wigner-2}
(\Delta x_{\rm canon.})_{\rm min} &\simeq& \sqrt{\frac{\hbar t}{2m}} = \frac{1}{\sqrt{2}}\sqrt{\lambda_{\rm C} r} \, ,
\notag \\
(\Delta v)_{\rm max} &\simeq& \sqrt{\frac{2\hbar}{m t}} = \sqrt{2}\sqrt{\frac{\lambda_{\rm C}}{r}}c \, ,
\end{eqnarray}
where we have again used $r = 2ct$. 

We note that similar arguments apply if we consider a modified experimental set up, in which a photon is emitted by the particle at $r=0$ and absorbed by a device in the lab frame at $r = ct$, or vice versa. 
(In other words, we note that reflection by the mirror is not an essential part of the experimental procedure and, in addition, that it does not affect the order of magnitude estimates of the minimum quantum uncertainty inherent in the measurement.)

For fundamental particles, it is therefore interesting to ask, what happens if a photon is emitted from the particle and {\it reabsorbed} within the interval $r \in (0,\lambda_{\rm C}]$? 
Strictly, the answer is that, for $r < \lambda_{\rm C}$, the non-relativistic theory breaks down and we must switch to a field theoretic picture. In this, the `measurement' of $r$ corresponds to a self-interaction in which the photon remains virtual. 
However, it is important to remember that interactions corresponding to measurements of $r < \Delta x_{\rm canon.}(r,m) < \lambda_{\rm C}(m)$ in the non-relativistic theory {\it are} physical. 
It is therefore reasonable to apply non-relativistic formulae, such as Eq. (\ref{Wigner-2}) and its gravitational `extensions', in this regime, on the understanding that `measuring' distances $r < \lambda_{\rm C}$ via photon emission / reabsorption corresponds to virtual, rather than real, photon exchange. 

A related point concerns the existence of superluminal velocities for $r \lesssim \lambda_{\rm C}$, as implied by Eq. (\ref{Wigner-2}). 
However, though virtual particles {\it can} travel faster than the speed of light, this does not imply a violation of causality, as information is not transmitted outside the light cone of a given space-time point \cite{Peskin:1995ev}. 
In fact, a similar effect occurs with respect to the standard Heisenberg term: for $\Delta x \lesssim  \lambda_{\rm C}$, the HUP implies $\Delta p \gtrsim mc$, or equivalently $\Delta v \gtrsim c$. 
Hence, superluminal velocities and sub-Compton probe distances in the non-relativistic theory are associated with the regime in which field theoretic effects become important. 
Nonetheless, we may continue to apply the non-relativistic formulae in this region, subject to the caveats stated above. 
These issues are discussed in detail in the Appendix. 

It is straightforward to extend the arguments presented in \cite{Calmet:2004mp,Calmet:2005mh} and \cite{Salecker:1957be} to include an estimate of the uncertainty in the position of the particle due to gravitational effects, $\Delta x_{\rm grav}$. 
By {\it assuming} that this is proportional to the Schwarzschild radius $r_{\rm S}$, Ng and van Dam defined the the total uncertainty due to canonical quantum effects, plus gravity, as 
\begin{eqnarray}  \label{dx_tot-2}
\Delta x_{\rm total}(r,m) &=& \Delta x_{\rm canon.}(r,m) + \Delta x_{\rm grav}(m)
\nonumber\\
&\gtrsim& \sqrt{\frac{\hbar r}{2mc}} + \beta\frac{Gm}{c^2} \, ,
\end{eqnarray}
where $\beta >0$, which is also assumed to be of order unity \cite{Ng:1993jb,Ng:1994zk}. 
(For $\beta = 2$, we recover $\Delta x_{\rm grav} = r_{\rm S}$ exactly.) 
Minimizing Eq. (\ref{dx_tot-2}) with respect to $m$ yields
\begin{eqnarray}  \label{M_min-1}
m \simeq \frac{1}{2}m_{\rm Pl}\left(\frac{r}{\beta^2l_{\rm Pl}}\right)^{1/3} \, , 
\end{eqnarray}
where $m_{\rm Pl} = \sqrt{\hbar c/G}$ is the Planck mass. 
Substituting this back into Eq. (\ref{dx_tot-2}), we obtain
\begin{eqnarray}  \label{MLUR-1}
\Delta x_{\rm total}(r,m) \geq (\Delta x_{\rm total})_{\rm min}(r) \simeq \frac{3}{2}(\beta l_{\rm Pl}^2r)^{1/3} \, .
\end{eqnarray} 
Neglecting numerical factors of order unity, and relabelling $\Delta s \rightarrow \Delta x_{\rm total}$ in Eq. (\ref{Karolyhazy_MLUR}), we see that Eq. (\ref{MLUR-1}) is equivalent to K{\' a}rolyh{\' a}zy's result with $r = ct \equiv s(t)$. 

\subsection{Motivations for the DE-UP} \label{sect2.4}

As shown above, in \cite{Karolyhazy:1966zz,KFL}, Eq. (\ref{Karolyhazy_MLUR}) was obtained by considering a gedanken experiment to measure the length of a space-time interval with minimum quantum uncertainty. 
This derivation relies on the fact that the mass of the measuring device (probe particle) $m$ distorts the background space-time. 
Equating the uncertainty in the particle's rest energy with uncertainty in its mass then implies an irremovable uncertainty or `haziness' in the space-time in the vicinity of the particle itself. 
This results in an absolute minimum uncertainty in the precision with which a \emph{gravitating} system can be used to measure the length of any given world-line, $s$. 
By contrast, the arguments presented in \cite{Ng:1993jb,Ng:1994zk} circumvent the need to assume quantum fluctuations in the rest mass, and hence the need to define a rest-energy Hamiltonian, $\hat{H}_{\rm rest} = \hat{m}c^2$.

Nonetheless, K{\' a}rolyh{\' a}zy's arguments \cite{Karolyhazy:1966zz,KFL} are similar to those of Ng and van Dam \cite{Ng:1993jb,Ng:1994zk}, in that $\beta \sim \mathcal{O}(1)$ arises as a direct result of the assumption that the Schwarzschild radius of a body, $r_{\rm S}(m) = 2Gm/c^2$, represents the minimum `gravitational uncertainty' in its position. 
In fact, for MLURs of the form (\ref{Karolyhazy_MLUR})/(\ref{MLUR-1}), it is usually assumed that $\beta \sim \mathcal{O}(1)$ in most of the existing quantum gravity literature \cite{Hossenfelder:2012jw,Garay:1994en}. 
For {\it all} the scenarios leading to Eq. (\ref{MLUR-1}), considered above, this is directly equivalent to assuming a minimum gravitational uncertainty of order $r_{\rm S}(m)$.

An important physical consequence is that, since Eq. (\ref{MLUR-1}) holds if and only if Eq. (\ref{M_min-1}) {\it also} holds, it is straightforward to verify
\begin{eqnarray}  \label{KFL-3}
(\Delta x_{\rm total})_{\rm min} \lessgtr \lambda_{\rm C} \iff m \lessgtr \frac{2}{3}m_{\rm Pl}\left(\frac{l_{\rm Pl}}{\beta r}\right)^{1/3} \, .
\end{eqnarray}
Substituting the minimization condition for $\Delta x_{\rm total}(r)$, Eq. (\ref{M_min-1}), into Eq. (\ref{KFL-3}) then gives
\begin{eqnarray}  \label{KFL-4}
r \lessgtr \frac{8}{3}\sqrt{\beta}l_{\rm Pl} \, .
\end{eqnarray}
For $\beta \sim \mathcal{O}(1)$, we require the `$>$' inequality in Eq. (\ref{KFL-4}), since many arguments imply that $l_{\rm Pl}$ represents the minimum resolvable length-scale due to quantum gravitational effects \cite{Hossenfelder:2012jw, Garay:1994en} . 
This implies that the `$>$' inequality also holds in Eq. (\ref{KFL-3}) and, hence, that the minimum quantum gravitational uncertainty predicted by K{\' a}rolyh{\' a}zy / Ng and van Dam is {\it always} greater than the Compton wavelength of the particle that minimizes it. 
Empirically, this poses a severe problem for the model, since since no such super-Compton limit has been observed for fundamental particles. 

However, physically, the assumption $\Delta x_{\rm grav}(m) \simeq r_{\rm S}(m)$ may be questioned on at least two grounds. 
First, we see that, for fundamental particles with masses $m \lesssim m_{\rm Pl}$, $\Delta x_{\rm grav}(m) \simeq r_{\rm S}(m) \lesssim l_{\rm Pl}$. 
Although the total uncertainty may remain super-Planckian, the assumption of simple additivity, $\Delta x_{\rm total}(r,m) = \Delta x_{\rm canon.}(r,m) + \Delta x_{\rm grav}(m)$, on which Eq. (\ref{MLUR-1}) is ultimately based, implies that canonical quantum uncertainty and the gravitational uncertainty arise {\it independently}, without influencing one other (i.e., that the gravitational uncertainty remains fixed, regardless of how dispersed the quantum wave packet becomes). 
It is therefore not clear whether a gravitational uncertainty given by $\Delta x_{\rm grav}(m) \simeq r_{\rm S}(m) < l_{\rm Pl}$ is physically meaningful. 
Second, gravity is a long range force. 
Intuitively, we may expect that, however it is defined, the gravitational uncertainty induced by the presence of a point-like or near point-like particle at $r = 0$ should fall with the gravtational field strength. 
Na{\" i}vely, we may assume that the gravitational uncertainty varies in proportion to the classical Newtonian potential, $\Delta x_{\rm grav}(r,m) \propto |\Phi(r,m)| \propto \beta(r) r_{\rm S}(m)$ as $r \rightarrow \infty$.  

If this is indeed the case, we see that, rather than being a simple constant, $\beta(r)$ must take the form of a ratio, $\beta(r) = \beta' l_{*}/r$, where $\beta' \sim \mathcal{O}(1)$ and $l_{*} \gg l_{\rm Pl}$ is a phenomenologically significant length-scale which is well motivated by fundamental physical considerations. 
In the context of a dark energy Universe, it is clear that the de Sitter horizon, $l_{\rm dS} = \sqrt{3/\Lambda}$, fulfils this criterion. 
As we shall see, one consequence of this is that states for which $r > l_{\rm Pl}$ {\it and} $(\Delta x_{\rm total})_{\rm min} < \lambda_{\rm C}$ become possible, in contrast to the predictions obtained from Eqs. (\ref{dx_tot-2})-(\ref{MLUR-1}). 
We also note that replacing $\beta = {\rm const.} \rightarrow \beta(r) = \beta' l_{*}/r$ in Eq. (\ref{dx_tot-2}) allows us to minimize $(\Delta x_{\rm total})_{\rm min}(r,m)$ with respect to either $m$ or $r$. 
It is straightforward to demonstrate that this minimum is unique, and is independent of both $r$ and $m$. 
As a result, the minimization procedure remains self-consistent in the limit $r \rightarrow \lambda_{\rm C}^{\pm}$. 
By contrast, rewriting $(\Delta x_{\rm total})_{\rm min}$ in Eq. (\ref{MLUR-1}) as a function of $m$ using Eq. (\ref{M_min-1}), then performing minimization with respect to $m$, we obtain a different result than if the minimization is performed with respect to $r$.

In Sec. \ref{sect3}, we derive MLURs in which the minimum uncertainty in a physical quantity $Q$ is given by the cube root of of three (possibly distinct) scales, $Q_1$, $Q_2$, $Q_3$, but which differ from relations derived from Eqs. (\ref{Karolyhazy_MLUR})/(\ref{MLUR-1}) in two important ways. 
First, the new relations attempt to incorporate the effects of dark energy, in the form of a cosmological constant, on the `smearing' of space-time and, thus, on the minimum quantum gravitational uncertainty inherent in a measurement of position. 
Second, they lead to substantially different but physically reasonable predictions in a number of scenarios. 
Specifically, they may be combined with other results obtained in general relativity and canonical quantum theory to give estimates of both the electron ($e^{-}$) and electron neutrino ($\nu_{e}$) masses, in terms of fundamental constants. 
These estimates yield the correct order of magnitude values obtained from experiment. 

In deriving the new relations we follow a procedure analogous to that used by Ng and van Dam \cite{Ng:1993jb,Ng:1994zk} (Sec. \ref{sect2.2.2}) but assume the existence of an asymptotically de Sitter / FLRW, rather than Minkowski, space-time. 
The results are obtained in two different ways. 
In the first, it is unnecessary to assume fluctuations in basic parameters, such as the mass $m$. This avoids the need to promote parameters to observables, represented by Hermitian operators in the non-relativistic quantum-gravitational regime. 
(From a technical point of view, it removes the need to define the operator $\hat{m}$ or, equivalently, the rest Hamiltonian $\hat{H}_{\rm rest} = \hat{m}/c^2$.) 
In this case, it is, however, necessary to make certain assumptions about the properties of space-time superpositions in the Newtonian limit. 
In particular, we {\it assume} the existence of an {\it upper bound} on $\Delta x_{\rm grav}$, given by the difference between line-elements in two classical space-times: one in which the particle is present and one in which it is absent. 
This is equivalent to assuming that the `spread' of quantum states $\Delta s$ cannot exceed the difference between the two classical extremes $|s'-s|$. 

In the second, we promote the classical Newtonian potential to an operator, $\Phi = -Gm/r \rightarrow \hat{\Phi} = -G\widehat{m/r}$, {\it {\` a} la} K{\' a}rolyh{\' a}zy, and estimate the associated uncertainty, $\Delta \Phi$, by considering a superposition of position states. 
We then relate $\Delta \Phi$ to $\Delta x_{\rm grav}$ by considering the associated uncertainty induced in the measurement of space-time intervals. 
From here on, we refer to all minimum quantity uncertainty relations of the form $(\Delta Q)_{\rm min} \simeq (Q_1Q_2Q_3)^{1/3}$ as `cubic', due to the value of the exponent on the right-hand side.  

\section{Dark energy-induced modifications of the MLUR -- the DE-UP} \label{sect3}

\subsection{Space-time uncertainty and classical perturbations -- a connection?} \label{sect3.1}

Like K{\' a}rolyh{\' a}zy, for $\tau \simeq \tau_0$, we take Eq. (\ref{Karolyhazy_s'}) as our starting point for the quantum mechanical definition of a `hazy' space-time. 
In this case, the Hubble flow correction term in Eq. (\ref{FLRW_pert_static-2}) is subdominant within the turn-around radius, $r \leq r_{\rm grav}$. 
However, rather than following the steps expressed in Eqs. (\ref{dm_K-1})-(\ref{dm_K-2}), leading to Eq. (\ref{Karolyhazy_MLUR}), we instead make the following physical assumption.

{\it We assume that the quantum mechanical uncertainty in the space-like interval between a particle of mass $m$ (located at $r=0$) and the coordinate distance $r$, is of the order of the difference between the classical values $s'(r,m)$ and $s(r)$, where $s(r) = s'(r,m)|_{m=0}$.}

Classically, the presence of the particle induces a perturbation in the background space-time, whose magnitude at $r$ is given by 
\begin{eqnarray} \label{ds_pert}
\Delta s_{\rm pert}(r,m) = |s'(r,m)-s(r)| \, ,
\end{eqnarray}
so that our assumption is equivalent to setting
\begin{eqnarray} \label{ds_assumption}
\Delta s(r,m) \simeq \Delta s_{\rm pert}(r,m) = |s'(r,m)-s(r)| \, ,
\end{eqnarray}
where $s(r)$ and $s'(r,m)$ represent the two classical extremes. 

In the classical picture, the underlying space-time may be in one of two {\it distinct} states. 
In the first, in which the particle is absent, the underlying metric corresponds to the unperturbed line element $s(r)$. In the second, in which the particle is present, the metric corresponds, instead, to the perturbed line element $s'(r,m)$. 
It is reasonable to suppose that, whatever the final theory of quantum gravity may be, a wave function of the form 
\begin{eqnarray} \label{st_wavefunction-1}
\ket{\Psi} = a \ket{s} + a' \ket{s'} \, ,  
\end{eqnarray}
describing a superposition of space-time background states, is possible in at least some limiting cases. 
Here, we use the notation $\Psi(t,\vec{r})$ to distinguish between wave functions representing space-time superpositions and $\psi(t,\vec{r})$, which represents a canonical quantum wave function that exists on a {\it definite} classical background. 
Though the mathematical formalism of a theory that contains both is not developed in the present work, we have in mind a composite wave function, that reduces to $\psi(t,\vec{r})$ when $\Psi(t,\vec{r})$ corresponds to a particular geometry.

More realistically, we may assume that the space-time background on which the canonical quantum wave function $\ket{\psi}$ propagates is, in fact, in a superposition of an infinite number of states, each corresponding to a unique classical line element $s$, i.e.
\begin{eqnarray} \label{psi_ST}
\ket{\Psi} = \int_{s_{\rm i}}^{s_{\rm f}} a(s)\ket{s} ds \, .
\end{eqnarray}
An expansion of this form will yield Eq. (\ref{ds_assumption}) if either the limits of integration are such that $s_{\rm i} = s$, the unperturbed line element, and $s_{\rm f} = s'$, the perturbed line element, or, more generally, if $s_{\rm i}$ and $s_{\rm f}$ take arbitrary values but the wave packet $\ket{\Psi}$ maintains a standard deviation of order $|s'-s|$. 
This holds true even for $s_{\rm i} \rightarrow 0$, $s_{\rm f} \rightarrow \infty$, as $s_{\rm i}$ and $s_{\rm f}$ reach the extremal classical limits.

Though a complete theory still eludes us, we may imagine a path integral over some kind of phase space, in which space-times corresponding to {\it all} other possible line-elements contribute negligible amplitudes to the total state vector expansion. 
These would include states corresponding to flat or negative curvature in the presence of $m$, as well as states giving rise to extreme positive curvature, which could only be sourced {\it classically} by much larger masses. 
This scenario is illustrated graphically in Fig. 3. 

\begin{figure}[h]
      \label{fig4}\centering
     \includegraphics[width=8.7cm]{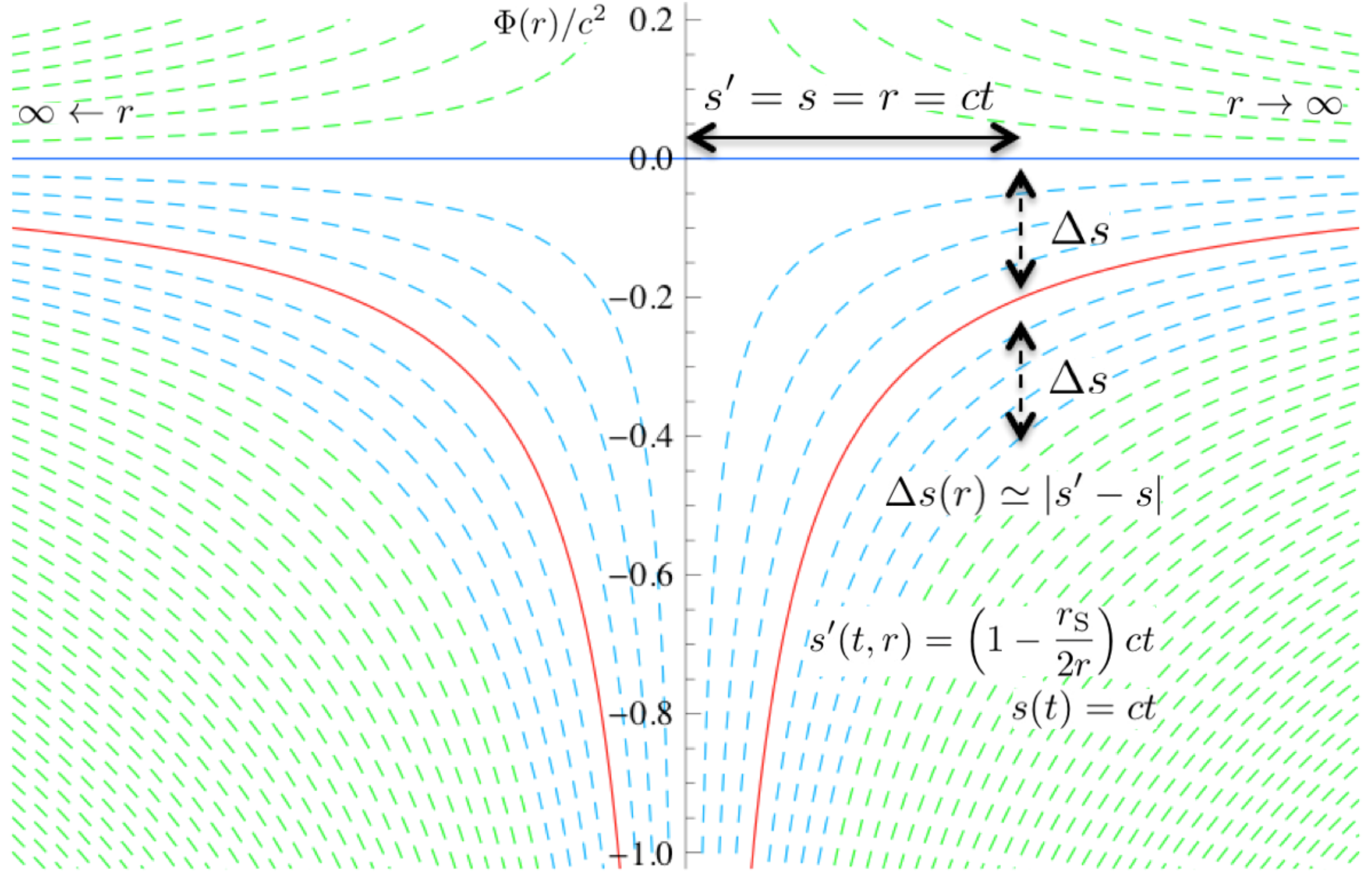}
     \caption{
     We assume that, quantum mechanically, space-time exists in a superposition of states. 
     If, {\it classically}, the presence of a particle of mass $m$ at $r=0$ induces a perturbation in the background space-time, of magnitude $\Delta s_{\rm pert}(r,m) = |s'(r,m) - s(r)|$, the spread of space-time states in the quantum superposition     
     is of order $\Delta s(r,m) \simeq \Delta s_{\rm pert}(r,m)$. 
     In this picture, the solid blue line represents the unperturbed geometry $s(r)$ and the solid red line represents the perturbed geometry $s'(r,m)$. Dashed blue lines represent states in the quantum superposition with non-negligible 
     amplitudes, whereas dashed green lines represent highly improbable states. 
     Typically, these correspond to flat, or even negative curvature in the presence of $m$ (upper half-plane), or to extreme positive curvature that could only be sourced {\it classically} by much higher masses (lower half-plane).}
\end{figure}

Incorporating the effects of universal expansion, we have $\Delta s(r,m) \rightarrow \Delta s(\tau,r,m)$.
Note that $\Delta s$ becomes a function of the cosmic time $\tau$ even if we choose to neglect the subdominant Hubble flow term in the perturbed Newtonian potential, since we must still shift to comoving coordinates $r \rightarrow a(\tau)r$. 
As described in Sec. \ref{sect2.2}, we may use a general two-part measurement procedure to resolve space-like intervals up to and including the particle horizon $\sim {\cal H}(\tau)/c$. 
In this case, the space-time uncertainty takes the form 
\begin{eqnarray} \label{dS_alt-2}
\Delta s(\tau,r,m) \simeq \frac{Gm}{c^2a(\tau)r}\frac{{\cal H}(\tau)}{c} \, .
\end{eqnarray}

With Eq. (\ref{dS_alt-2}) as our new starting point, we may now ask the question: how is this scenario affected by the presence of dark energy, in the form of a cosmological constant $\Lambda$? 
Clearly, the main physical consequence at the present epoch ($\tau \simeq \tau_0$, $a \simeq a_0 = 1$) is the existence of a cosmological horizon at a {\it fixed} distance from any observer for all $\tau \gtrsim \tau_0$. 
This is the de Sitter horizon, which corresponds to the unperturbed space-like interval $s(\tau_0) \simeq \mathcal{H}_0/c \simeq l_{\rm dS} = \sqrt{3/\Lambda}$. 
Therefore: 
\begin{eqnarray} \label{dS_alt-2*}
\Delta s(\tau_0,r,m) \simeq \frac{Gm}{c^2r} \, l_{\rm dS} \, .
\end{eqnarray}

Thus, in applying Eq. (\ref{dS_alt-2*}) to particles at the present epoch, we have in mind a particle interacting {\it simultaneously} with an object at $r$, close to its CoM, and with the furthest reaches of its environment, represented by ${\cal H}_0/c \simeq l_{\rm dS}$. For $r > \lambda_{\rm C}$, this object may be a detector in the lab frame, which simultaneously receives signals (e.g. photons) from the particle and from distant objects close to $l_{\rm dS}$. 
However, for $r < \lambda_{\rm C}$, the local object with which the probe particle interacts is simply itself and the local interaction involves the exchange of virtual particles. 
In principle, the long-range interaction between $r < \lambda_{\rm C}$ and $l_{\rm dS}$ may also involve the exchange of virtual, as well as real, particles. 

For our purposes, the fact that the interaction between the particle and its horizon may involve the exchange of virtual particles is extremely important. 
In effect, such interactions constantly `measure' the distance from the particle's CoM -- or, more specifically, from a point $r < \lambda_{\rm C}$ close to the CoM -- to its horizon. 
Hence, any irremovable uncertainty present in the result of this measurement is {\it equivalent} to an irremovable uncertainty in the position of the particle. 
Classically, both the position of the CoM and the position of the horizon are well defined, so that any quantum uncertainty in the distance between them is equivalent to an uncertainty in the position of either (or both).  

Next, we note that we may obtain the same result, Eq. (\ref{dS_alt-2*}), using an operational procedure in canonical QM. 
The derivation proceeds as follows. 
In the classical picture, a point-particle of mass $m$, located at $\vec{r}'$ (i.e., represented by the function $\delta(\vec{r}-\vec{r}'$)), generates a well-defined gravitational potential at a general point $\vec{r}$, given by 
\begin{eqnarray} \label{Phi_classical}
\Phi(m,|\vec{r}-\vec{r}'|) = -\frac{Gm}{|\vec{r}-\vec{r}'|}\ \, . 
\end{eqnarray}
In the quantum picture, the classical potential is promoted to an operator, $\Phi \rightarrow \hat{\Phi}$, such that
\begin{eqnarray} \label{Phi_Op}
\hat{\Phi}(m,|\vec{r}-\vec{r}'|)\delta(\vec{r}-\vec{r}') \equiv -\frac{Gm}{|\vec{r}-\vec{r}'|}\delta(\vec{r}-\vec{r}') \, . 
\end{eqnarray}
In other words, acting on the canonical position eigenstate $\delta(\vec{r}-\vec{r}')$, $\hat{\Phi}$ recovers the classical potential $\Phi$. 
For superpositions of position states, $\psi(t,\vec{r}')$, the gravitational potential at $\vec{r}$ will also be given by a superposition of states. We then have $\langle \hat{\Phi} \rangle = \langle \psi | \hat{\Phi} | \psi \rangle$, yielding
\begin{eqnarray}
\langle \hat{\Phi} \rangle = -\int \frac{Gm}{|\vec{r}-\vec{r}'|}|\psi(t,\vec{r}')|^2 d^3r' \, ,
\end{eqnarray}
and higher-order moments $\langle \hat{\Phi}^n \rangle$ may be defined in like manner. 
In the limit $|\psi(t,\vec{r}')|^2 \rightarrow \delta(\vec{r} - \vec{r}')$ the classical potential is recovered. 

As a concrete example, we consider spherically symmetric Gaussian states, for which
\begin{equation}  \label{Gaussian-1}
\Delta \Phi = \sqrt{\langle \hat{\Phi}^2 \rangle - \langle \hat{\Phi} \rangle^2} \simeq 
\left \lbrace
\begin{array}{rl}
&-\frac{Gmr}{(\Delta x)^2} \quad  \ (r \lesssim \Delta x) \\
&-\frac{Gm\Delta x}{r^2} \quad (r \gtrsim \Delta x) 
\end{array}
\right.
\end{equation}
where we have chosen our coordinate system so that the wave packet CoM is located at $\vec{r} = 0$ and $\Delta x$ denotes the canonical quantum uncertainty. 
For Gaussian states, this is given by
\begin{eqnarray}
&&\Delta x(t) = \sigma \left(1 + \frac{\hbar^2 t^2}{m^2\sigma^4} \right)^{1/2} 
\nonumber\\
&\equiv& \Delta x(r) = \sigma \left(1 + \frac{r^2\lambda_{\rm C}^2}{\sigma^4} \right)^{1/2} \, , 
\end{eqnarray}
where $\sigma = \Delta x(0)$ is the initial spread at $t=0$ and $r \equiv ct$. 

For $\sigma \gtrsim (\Delta x_{\rm canon.})_{\rm min} \simeq \sqrt{\lambda_{\rm C}r}$ ($r \lesssim \sigma^2/\lambda_{\rm C}$), the spread of the wave function is given approximately by $\Delta x(r) \simeq \sigma$, whereas, for $\sigma \lesssim \sqrt{\lambda_{\rm C}r}$ ($r \gtrsim \sigma^2/\lambda_{\rm C}$), the late-time spread is given by $\Delta x(r) \simeq \lambda_{\rm C}r/\sigma$. 
Hence, any `measurements' (including self-interactions) occurring on time-scales $t \lesssim \sigma^2/(c\lambda_{\rm C})$ are effectively instantaneous and do not significantly disturb the initial ($t=0$) quantum state. 
However, as $t=0$ is an idealization, which is likely not physically realizable, we restrict our attention henceforth to time-scales $t \gtrsim \sigma^2/(c\lambda_{\rm C})$. 
Eq. (\ref{Gaussian-1}) then gives
\begin{equation} \label{Gaussian-2}
\Delta \Phi \simeq 
\left \lbrace
\begin{array}{rl}
&-\frac{Gm\sigma^2}{\lambda_{\rm C}^2r} \quad \ (\sigma \lesssim \lambda_{\rm C}) \\
&-\frac{Gm\lambda_{\rm C}}{\sigma r} \quad \ (\sigma \gtrsim \lambda_{\rm C}) \, . 
\end{array}
\right.
\end{equation}
The two expressions above coincide for $\sigma \simeq \lambda_{\rm C}$ -- which is a reasonable assumption in the canonical theory -- yielding
\begin{eqnarray} \label{Gaussian-3}
\Delta \Phi \simeq |\Phi(m,r) - \Phi(0,r)| \simeq -\frac{Gm}{r} \, .
\end{eqnarray}

The next step is to determine the relationship between $\Delta \Phi$ and $\Delta s$, the uncertainty in the measured space-time interval. 
This can be done by setting $\tau \simeq \tau_0$ in Eq. (\ref{FLRW_pert_static-2}) and ignoring the sub-dominant Hubble flow term for $r \leq r_{\rm grav}$, giving 
\begin{eqnarray} \label{class_pert}
s'(\tau_0) \simeq \left(1 - \frac{\Phi(m,r)}{c^2}\right) s(\tau_0) \, , 
\end{eqnarray}
where $s(\tau_0) \simeq l_{\rm dS} = {\rm const.}$. In the quantum picture, we then have
\begin{eqnarray} \label{quant_pert}
\hat{s}'(\tau_0) \simeq \left(1 - \frac{\hat{\Phi}(m,r)}{c^2}\right) l_{\rm dS} \, , 
\end{eqnarray}
giving
\begin{eqnarray}
\Delta s'(r,m) &\simeq& \frac{\Delta \Phi(r,m)}{c^2} l_{\rm ds} 
\notag\\
&\simeq& \frac{|\Phi(m,r) - \Phi(0,r)|}{c^2} l_{\rm ds} 
\notag\\
&\simeq& |s' - s| \, , 
\end{eqnarray}
as claimed. $\Delta s'(r,m)$ can then be identified with $\Delta x_{\rm grav}$, as before. 

Note that here, as in the derivations of the canonical quantum uncertainty given in \cite{Calmet:2004mp,Calmet:2005mh} and \cite{Salecker:1957be}, we continue to identify $r = ct$ in deriving the expression for $\Delta \Phi$. In this sense, the geometric nature of the gravitational field is not explicitly accounted for in this step and (\ref{Phi_classical}) is treated like any other potential existing {\it on} a flat-space background.  
This is an unavoidable limitation of working {\it within} the framework of canonical QM up to this point. 

However, combining this with the classical relation given by Eqs. (\ref{FLRW_pert_static-2})/(\ref{class_pert}), and `quantizing' the latter by promoting the classical potential to an operator $\Phi \rightarrow \hat{\Phi}$, allows us to obtain an expression for the standard deviation of the space-time interval operator $\hat{s}'$. 
This takes us beyond canonical QM. 
Thereafter, the geometric nature of $\hat{\Phi}$ is made explicit -- via its relation to $\hat{s}'$ -- and the $r$ appearing in Eq. (\ref{quant_pert}) {\it cannot} be identified with the flat-space interval corresponding to the unperturbed line-element, i.e. $r \neq s(t) = ct$. 
Nonetheless, it is interesting to note that such a procedure yields results analogous to Eq. (\ref{ds_assumption}), which is explicitly based on the physical picture illustrated in Fig. 3. 
At the very least, we may say that this picture does not contradict the results of canonical QM, but allows us to reinterpret a canonically quantized $\Phi$ in terms of a K{\' a}rolyh{\' a}zy-type `hazy' space-time. 

\subsection{Derivation of the DE-UP} \label{sect3.2}

Since, at the present epoch, the particle's communication with the outside world is confined within the de Sitter radius \cite{Spradlin:2001pw} -- that is, within the region $r \in [0,l_{\rm dS})$ -- the minimum value of the gravitational uncertainty, induced at a given point $r$ from its CoM, is 
\begin{eqnarray} \label{dS_alt-1}
\Delta x_{\rm grav}(r,m) \simeq \frac{Gm}{c^2r}l_{\rm dS} \, .
\end{eqnarray}
This is simply Eq. (\ref{dS_alt-2*}) with $\Delta s \equiv \Delta x_{\rm grav}$.
To this we must add the canonical uncertainty due to the gradual diffusion of the wave function, predicted by canonical non-gravitational QM. 
We here assume that the respective uncertainties are additive, which is consistent with the perturbative approach to the gravitational sector, considered in Sec. \ref{sect2}. 
We then have
\begin{eqnarray} \label{Total_Uncertainty-2}
\Delta x_{\rm total}(\Delta v,r,m) &=& \Delta x_{\rm canon.}(\Delta v,r,m) + \Delta x_{\rm grav}(r,m)
\nonumber\\ 
&\geq& \Delta x(\Delta v) + \Delta x_{\rm recoil}(\Delta v,r,m) 
\nonumber\\ 
&+& \Delta x_{\rm grav}(r,m) 
\nonumber\\ 
&\geq&  (\Delta x_{\rm canon.})_{\rm min}(r,m) 
\nonumber\\ 
&+& \Delta x_{\rm grav}(r,m) \, .
\end{eqnarray}

Instead of using the order of magnitude estimates for $\Delta x_{\rm recoil}$ and $\Delta x_{\rm grav}$, obtained in Secs. \ref{sect2.4} and \ref{sect3.1}-\ref{sect3.2}, together with an order of magnitude inequality `$\gtrsim$', we assume that Eq. (\ref{Total_Uncertainty-2}) holds exactly when these quantities are defined precisely, up to appropriate numerical factors. 
Hence, we introduce two numerical constants, $\alpha', \beta' > 0$, which are assumed to be not hierarchically different to unity. 
Eq. (\ref{Total_Uncertainty-2}) may then be rewritten as
\begin{eqnarray} \label{DE-UP-1}
\Delta x_{\rm total}(\Delta v,r,m) \geq \frac{\lambda_{\rm C}}{2}\frac{c}{\Delta v} + \alpha' \frac{\Delta v}{c}r + \beta' \frac{l_{\rm Pl}^2l_{\rm dS}}{\lambda_{\rm C}r} \, ,
\end{eqnarray}
where $\alpha', \beta' \sim \mathcal{O}(1)$. 
Note that, since $\alpha', \beta'$ are required to be of order unity, they do not represent free parameters of the model. 
If the value of either constant were permitted to be hierarchically larger or smaller than one, this would alter the existing mass / length-scales present in the theory, indicating new physics. 
The physical basis of Eq. (\ref{DE-UP-1}), considered in the preceding sections, does not allow for this. 
From here on, we refer to Eq. (\ref{DE-UP-1}) as the DE-UP-1. 

Minimizing the DE-UP-1 with respect to $\Delta v$ yields
\begin{eqnarray} \label{DE-UP-1-mins}
(\Delta x_{\rm canon.})_{\rm min} = \sqrt{2\alpha'}\sqrt{\lambda_{\rm C} r} \, ,
\nonumber\\
(\Delta v)_{\rm max} = \frac{1}{\sqrt{2\alpha'}}\sqrt{\frac{\lambda_{\rm C}}{r}}c \, ,
\end{eqnarray}
and hence
\begin{eqnarray} \label{DE-UP-2}
\Delta x_{\rm total}(r,m) \geq \sqrt{2\alpha'}\sqrt{\lambda_{\rm C} r} + \beta'\frac{l_{\rm Pl}^2l_{\rm dS}}{\lambda_{\rm C}r} \, .
\end{eqnarray}
From here on, we refer to Eq. (\ref{DE-UP-2}) as the DE-UP-2. 

Finally, minimizing the DE-UP-2 with respect to $m$ or $r$, gives 
\begin{eqnarray} \label{DE-UP-2-mins}
&& m = \left(\frac{\alpha'}{2\beta'^2}\right)^{1/3} \frac{r}{(l_{\rm Pl}^2l_{\rm dS})^{1/3}} (m_{\rm Pl}^2m_{\rm dS})^{1/3} 
\nonumber\\
&\iff& r \equiv r_{\rm min} \equiv \left(\frac{2\beta'^2}{\alpha'}\right)^{1/3}(l_{\rm Pl}l_{\rm dS}^2)^{1/3}\frac{m}{m_{\rm Pl}} \, ,
\end{eqnarray}
where $r_{\rm min}$ denotes the value of the probe distance $r$ that minimizes the total uncertainty. 
Hence,
\begin{eqnarray} \label{DE-UP-2-mins-A}
\Delta x_{\rm grav} = \left(\frac{\alpha'\beta'}{2}\right)^{1/3} (l_{\rm Pl}^2l_{\rm dS})^{1/3} \, , 
\end{eqnarray}
and
\begin{eqnarray} \label{DE-UP-2-mins-B}
(\Delta x_{\rm canon.})_{\rm min} &=& (4\alpha'\beta')^{1/3} (l_{\rm Pl}^2l_{\rm dS})^{1/3} \, , 
\nonumber\\
(\Delta v)_{\rm max} &=& \left(\frac{1}{4\alpha'\beta'}\right)^{1/3}\frac{\lambda_{\rm C}}{(l_{\rm Pl}^2l_{\rm dS})^{1/3}} c \, ,
\end{eqnarray}
so that
\begin{eqnarray} \label{DE-UP-3}
(\Delta x_{\rm total})_{\rm min} = 3\left(\frac{\alpha'\beta'}{2}\right)^{1/3}(l_{\rm Pl}^2l_{\rm dS})^{1/3} \, .
\end{eqnarray}
From here on, we refer to Eq. (\ref{DE-UP-3}) as the DE-UP-3.
 
It is straightforward to show that minimizing the DE-UP-1 with respect $r$, followed by $m$ {\it or} $\Delta v$, or with respect to $m$, followed by $\Delta v$ {\it or} $r$, yields the same final result (\ref{DE-UP-3}). 
Viewed as a function of all three variables, $\Delta x_{\rm total}(\Delta v,r,m)$ has a {\it unique} minimum. 

After completely minimizing $\Delta x_{\rm total}(\Delta v,r,m)$ to obtain the DE-UP-3 (\ref{DE-UP-3}), an interesting critical mass-scale is obtained by setting the recoil velocity of the particle equal to the speed of light,
\begin{eqnarray} \label{critical_mass}
&&(\Delta v)_{\rm max} = c
\nonumber\\
&\iff& \lambda_{\rm C} = (\Delta x_{\rm canon.})_{\rm min} = (4\alpha'\beta')^{1/3} (l_{\rm Pl}^2l_{\rm dS})^{1/3} 
\nonumber\\
&\iff& m = m_{\rm crit.} \equiv \left(\frac{1}{4\alpha'\beta'}\right)^{1/3}  (m_{\rm Pl}^2m_{\rm dS})^{1/3} \, .
\end{eqnarray}
The unique properties of this mass, including its relevance for holography in an asymptotically de Sitter Universe, were considered in \cite{Burikham:2016rbj}. 
In addition, we note that all expressions, Eqs. (\ref{DE-UP-1})-(\ref{DE-UP-3}), are invariant under simultaneous re-scalings of the form
\begin{eqnarray} \label{rescalings}
\Delta v &\rightarrow& \alpha_{\rm Q}^{-1}\Delta v \, ,
\nonumber\\
m &\rightarrow& \alpha_{\rm Q}m \, ,
\nonumber\\
r &\rightarrow& \alpha_{\rm Q}r \, ,
\end{eqnarray}
where $\alpha_{\rm Q} > 0$ is a positive real parameter, which does not depend on any of the three variables $\Delta v$, $m$, or $r$. 
In Sec. \ref{sect3.3.2}, we will show that $\alpha_{\rm Q}$ may depend, at most, on the charge $Q$ of the probe particle. 
Eq. (\ref{critical_mass}) then becomes
\begin{eqnarray} \label{critical_mass_rescale}
&&(\Delta v)_{\rm max} = \alpha_{\rm Q}^{-1}c
\nonumber\\
&\iff& \lambda_{\rm C} = \alpha_{\rm Q}^{-1}(\Delta x_{\rm canon.})_{\rm min} = \alpha_{\rm Q}^{-1}(4\alpha'\beta')^{1/3}(l_{\rm Pl}^2l_{\rm dS})^{1/3} 
\nonumber\\
&\iff& m = \alpha_{\rm Q}m_{\rm crit.} = \alpha_{\rm Q} \left(\frac{1}{4\alpha'\beta'}\right)^{1/3} (m_{\rm Pl}^2m_{\rm dS})^{1/3} \, .
\end{eqnarray}

In summary, the first two terms in the DE-UP-1 (\ref{DE-UP-1}), $\Delta x$ and $\Delta x_{\rm recoil}$, give the canonical quantum uncertainty inherent in the measurement of a distance $r$. 
This distance is measured by means of a force-mediating boson emitted from a probe particle of mass $m$, whose CoM is initially located at $r=0$, and its subsequent absorption by a `detector' at $r>0$. For $r \in (0,\lambda_{\rm C}]$, the detector is simply the particle itself, and the boson remains {\it virtual}. 
This uncertainty is {\it equivalent} to the canonical uncertainty in the position of the particle, as viewed by an observer situated at $r$. 
For charged particles, the relevant boson is a photon but, for neutral particles, it may instead be a short-range massive boson. 
We note that, in the canonical non-gravitational theory, the distance $r$ and the time taken to perform the measurement $t$ are related via $r = ct$. 

The third term $\Delta x_{\rm grav}$ represents the gravitational uncertainty at $r$ due to the `haziness' of the underlying space-time metric, induced by the presence of the particle. 
This is equivalent to the irremovable uncertainty inherent in a measurement of the horizon distance ${\cal H}_0/c \simeq l_{\rm dS}$. 
The measurement is completed by means of real or {\it virtual} boson exchange between the probe particle and a `detector' at $r>0$, and between $l_{\rm dS}$ and $r$. 
As with the canonical uncertainty, for $r \in (0,\lambda_{\rm C}]$, the detector is simply the particle itself. 
In the non-relativistic picture, the space-time haziness is related to the haziness of the Newtonian potential, which exists in a superposition of states (\ref{Gaussian-3}). 
We note that, once gravitational effects are taken into account, the simple relationship between the time taken to perform the measurement $t$ and the coordinate distance $r$ no longer holds, $r \neq ct$. 

Together, all three terms give the total uncertainty, incorporating both the uncertainty in the space-time metric -- including the effects of dark energy in the form of a cosmological constant $\Lambda$ -- and the canonical uncertainty in the position of the particle's CoM. 

\subsection{Basic properties of the DE-UP} \label{sect3.3}

\subsubsection{Application to neutral particles} \label{sect3.3.1}

We now investigate the basic properties of the DE-UP. Since $l_{\rm Pl}$ is expected to form a fundamental lower bound on the resolvability of all physically measurable length-scales \cite{Garay:1994en}, we start by imposing the conditions $(\Delta x_{\rm canon.})_{\rm min}, \ \Delta x_{\rm grav}, \ r \geq l_{\rm Pl}$. 
As we show below, imposing all three constraints gives rise to a fundamental lower bound on the mass of a system obeying Eq. (\ref{DE-UP-2}). 
Furthermore, this bound may be derived independently by combining minimum-density requirements, obtained from the generalized Buchdahl inequalities for a spherically symmetric system in the presence of dark energy ($\Lambda > 0$) \cite{Buchdahl:1959zz,Mak:2001gg,Boehmer:2005sm,Boehmer:2006fd}, with the simple requirement of the existence of a Compton wavelength \cite{Burikham:2015nma}. 
Hence, beginning with the independently derived result, we see that both the canonical and gravitational terms in the DE-UP-2 (\ref{DE-UP-2}), together with the probe distance $r$, remain super-Planckian under physically reasonable conditions. 

Since we require $\Delta x_{\rm grav} \geq l_{\rm Pl}$, let us paramaterize it such that
\begin{eqnarray} \label{Delta_x_grav_param}
\Delta x_{\rm grav} = c_1 l_{\rm Pl} \, , \quad (c_1 \geq 1) \, ,
\end{eqnarray}
giving
\begin{eqnarray} \label{Delta_x_grav_param_r}
r = \frac{\beta'}{c_1}\frac{m}{m_{\rm Pl}} l_{\rm dS} \, .
\end{eqnarray}
Likewise, setting $(\Delta x_{\rm canon.})_{\rm min} \geq l_{\rm Pl}$ so that
\begin{eqnarray} \label{Delta_x_canon_param}
(\Delta x_{\rm canon.})_{\rm min} = c_2 l_{\rm Pl} \, , \quad (c_2 \geq 1) \, ,
\end{eqnarray}
gives
\begin{eqnarray} \label{Delta_x_canon_param_r}
r = \frac{c_2^2}{2\alpha'} \frac{m}{m_{\rm Pl}} l_{\rm Pl} \, .
\end{eqnarray}
For later convenience, we now define the ratio
\begin{eqnarray} \label{N}
N = \left(\frac{l_{\rm dS}}{l_{\rm Pl}}\right)^2 = \frac{3c^3}{\hbar G \Lambda} = 1.030 \times 10^{122} \, ,
\end{eqnarray}
and, comparing Eqs. (\ref{Delta_x_grav_param_r}) and (\ref{Delta_x_canon_param_r}), we have
\begin{eqnarray} \label{c_1c_2^2}
\frac{c_1c_2^2}{2\alpha'\beta'} = \frac{l_{\rm dS}}{l_{\rm Pl}} = N^{1/2} \, .
\end{eqnarray}
To within numerical factors of order unity, $N$ equals the number of Planck-sized bits on the present day boundary of the observable Universe or, equivalently, the number of cells with volume $\sim (\Delta x_{\rm total})_{\rm min}^3$ in the present day bulk \cite{Burikham:2015sro}. 

Let us also require $r \geq l_{\rm Pl}$, setting
\begin{eqnarray} \label{r_param}
r = c_3l_{\rm Pl} \, , \quad (c_3 \geq 1) \, .
\end{eqnarray}
Combining this with Eqs. (\ref{Delta_x_grav_param_r}) and (\ref{Delta_x_canon_param_r}) yields
\begin{eqnarray} \label{m_param-1}
m = \frac{c_1c_3}{\beta'}m_{\rm dS} 
\end{eqnarray}
and
\begin{eqnarray} \label{m_param-2}
m = \frac{2\alpha' c_3}{c_2^2}m_{\rm Pl} \, ,
\end{eqnarray}
respectively, which themselves combine to give
\begin{eqnarray} \label{m_param-2}
m = \sqrt{\frac{2\alpha'}{\beta'}}\frac{\sqrt{c_1}c_3}{c_2}\sqrt{m_{\rm Pl}m_{\rm dS}} \, .
\end{eqnarray}

We now define a new mass-scale, 
\begin{eqnarray} \label{m_Lambda}
m_{\Lambda} = \frac{1}{\sqrt{2}}\sqrt{m_{\rm Pl}m_{\rm dS}} = 4.832 \times 10^{-36} \rm g \, .
\end{eqnarray}
It is straightforward to demonstrate that $m_{\Lambda}$ is the minimum mass of a stable, spherically symmetric, gravitating, charge-neutral and quantum mechanical object. 
This result was first obtained in \cite{Burikham:2015nma}, though we briefly review its derivation for the sake of clarity. 

In \cite{Boehmer:2005sm}, it was shown that the density of a stable, spherically symmetric, gravitating, charge-neutral and {\it classical} compact object must satisfy the inequality 
\begin{eqnarray} \label{min_dens}
\rho \geq \rho_{\rm min} =  \rho_{\Lambda}/2 \, , 
\end{eqnarray}
where $\rho_{\Lambda}$ is the dark energy density given by Eq. (\ref{Lambda_E_p}). 
Though the proof of this statement, which follows directly from the generalised Buchdahl inequalities \cite{Buchdahl:1959zz,Mak:2001gg,Boehmer:2005sm,Boehmer:2006fd}, is rather complicated, its physical meaning is intuitively clear: compact objects with energy densities significantly lower than the vacuum density have insufficient self-gravity to overcome the repulsive effect of dark energy. 
For bodies of fixed mass $m$, classical radius $R$ and initial density $\rho < \rho_{\rm min}$, the spatial expansion caused by $\Lambda > 0$, which acts as repulsive force, causes $R$ to expand indefinitely and the object is unstable. 
For a quantum mechanical particle, whose mass $m$ is localized within a sphere of radius $R = \lambda_{\rm C}$, we then have 
\begin{eqnarray} \label{min_mass}
\rho = \frac{3}{4\pi}\frac{m^4c^3}{\hbar^3} \geq \rho_{\rm min}  = \frac{\Lambda c^2}{16 \pi G} \iff m \geq m_{\Lambda} \, .
\end{eqnarray}
Since $c_1,c_2,c_3 \geq 1$ by construction, we may then identify 
\begin{eqnarray} \label{alpha'_beta'_rel-1}
\beta' = 4\alpha' \, . 
\end{eqnarray}
For later convenience, we define 
\begin{eqnarray} \label{l_Lambda}
l_{\Lambda} = \sqrt{2} \sqrt{l_{\rm Pl}l_{\rm dS}} = 7.283 \times 10^{-3} {\rm cm} \, ,
\end{eqnarray}
which is simply the reduced Compton wavelength associated with the minimum mass $m_{\Lambda}$.

Thus, requiring both the individual components of the DE-UP-2 (\ref{DE-UP-2}) and the probe length $r$ to be super-Planckian ensures its consistency with both general relativistic and quantum mechanical constraints. 
These, in turn, allow us to fix the relation between the parameters $\alpha'$ and $\beta'$ on purely theoretical grounds. 
However, we must remember that, in reality, a length-scale of order $l_{\rm Pl}$, up to numerical factors of order unity, may be the true fundamental cut-off for resolvable length-scales in nature, so that this relation must be taken as tentative and some ambiguity still remains. 

Equivalently, we see that, beginning with the result $m \geq m_{\Lambda}$ and reversing our previous logic, the existence of a minimum stable mass for self-gravitating quantum mechanical objects ensures that all three length-scales $(\Delta x_{\rm canon.})_{\rm min}$, $\Delta x_{\rm grav}$ and $r$, appearing in the DE-UP-2 (\ref{DE-UP-2}), remain super-Planckian.  
We now investigate permissible ranges of each of these length-scales for masses in the range $m_{\Lambda} \leq m \leq 2^{-1/2}m_{\rm Pl}$, which corresponds to the fundamental particle regime.

Rearranging Eq. (\ref{c_1c_2^2}) and imposing $c_2 \geq 1$ yields
\begin{eqnarray} \label{c1_constraints}
1 \leq c_1 \leq 2\alpha'\beta' \frac{l_{\rm dS}}{l_{\rm Pl}} \, ,
\end{eqnarray}
while imposing $c_1 \geq 1$ gives
\begin{eqnarray} \label{c2_constraints}
1 \leq c_2 \leq \sqrt{2\alpha'\beta'} \sqrt{\frac{l_{\rm dS}}{l_{\rm Pl}}} \, .
\end{eqnarray}
Substituting (\ref{c2_constraints}) into (\ref{Delta_x_canon_param}) then gives
\begin{eqnarray} \label{}
l_{\rm Pl} \leq (\Delta x_{\rm canon.})_{\rm min}(r,m) \leq \sqrt{2\alpha'\beta'}\sqrt{l_{\rm Pl}l_{\rm dS}}\, ,
\end{eqnarray}
and
\begin{eqnarray} \label{r_limits-1}
l_{\rm Pl} \leq r \leq \beta'\frac{m}{m_{\rm Pl}}l_{\rm dS} \, ,
\end{eqnarray}
where the upper bound is equivalent to the condition $\Delta x_{\rm grav}(r,m) \geq l_{\rm Pl}$. 
Next, we impose the following condition, stemming from Eq. (\ref{Delta_x_grav_param_r}) with $c_1 \geq 1$:
\begin{eqnarray} \label{}
r \leq \beta'\frac{m}{m_{\rm Pl}}l_{\rm dS}  \leq  l_{\rm dS} \iff m  \leq \frac{m_{\rm Pl}}{\beta'} \, . 
\end{eqnarray}
Hence, setting
\begin{eqnarray} \label{alpha'beta'}
\beta'  = \sqrt{2} \iff \alpha' = \frac{1}{2\sqrt{2}}
\end{eqnarray}
allows us to recover the standard constraint
\begin{eqnarray} \label{standard_cons-1}
m  \leq  2^{-1/2}m_{\rm Pl} \iff \lambda_{\rm C} \geq r_{\rm S} \, ,
\end{eqnarray}
which defines the fundamental particle domain. 
We use the values $\alpha' = 1/(2\sqrt{2})$ and $\beta' = \sqrt{2}$ from here on, unless explicitly stated. 

Thus, for fundamental particles, the ranges of $m$ and $r$ are restricted such that
\begin{eqnarray} \label{m_r_limits}
m_{\Lambda} = \frac{1}{\sqrt{2}}\sqrt{m_{\rm Pl}m_{\rm dS}} \leq &m& \leq \frac{1}{\sqrt{2}}m_{\rm Pl} \, ,
\nonumber\\
l_{\rm Pl} \leq &r& \leq \sqrt{2}\frac{l_{\rm dS}l_{\rm Pl}}{\lambda_{\rm C}} \, ,
\end{eqnarray}
giving
\begin{eqnarray} \label{uncertainties_limits}
2^{-1/4}\sqrt{\lambda_{\rm C}l_{\rm Pl}} \leq (\Delta x_{\rm canon.})_{\rm min}  \leq \sqrt{l_{\rm Pl}l_{\rm dS}} \, ,
\nonumber\\
\sqrt{2}\frac{l_{\rm dS}l_{\rm Pl}}{\lambda_{\rm C}} \geq  \Delta x_{\rm grav} \geq l_{\rm Pl} \, .
\end{eqnarray}
For the limiting mass scales $m_{\Lambda}$ and $2^{-1/2}m_{\rm Pl}$, and the critical mass scale $m_{\rm crit}$ (\ref{critical_mass}), we have
\begin{eqnarray} \label{m_Lambda_limits}
m &=& m_{\Lambda}
\nonumber\\
&\iff& l_{\rm Pl} \leq r \leq \sqrt{l_{\rm Pl}l_{\rm dS}} \, ,
\nonumber\\
&&(l_{\rm Pl}^3l_{\rm dS})^{1/4} \leq (\Delta x_{\rm canon.})_{\rm min} \leq \sqrt{l_{\rm Pl}l_{\rm dS}} \, ,
\nonumber\\
&&\sqrt{l_{\rm Pl}l_{\rm dS}} \geq \Delta x_{\rm grav} \geq l_{\rm Pl} \, ,
\end{eqnarray}
\begin{eqnarray} \label{m_crit_limits}
m &=& m_{\rm crit}
\nonumber\\
&\iff& l_{\rm Pl} \leq r \leq 2^{1/6}(l_{\rm Pl}l_{\rm dS}^2)^{1/3} \, ,
\nonumber\\
&&2^{-1/12}(l_{\rm Pl}^5l_{\rm dS})^{1/6} \leq (\Delta x_{\rm canon.})_{\rm min} \leq \sqrt{l_{\rm Pl}l_{\rm dS}} \, ,
\nonumber\\
&&2^{1/6}(l_{\rm Pl}l_{\rm dS}^2)^{1/3} \geq \Delta x_{\rm grav} \geq l_{\rm Pl} \, .
\end{eqnarray}
and
\begin{eqnarray} \label{m_Planck_limits}
m &=& 2^{-1/2}m_{\rm Pl} 
\nonumber\\
&\iff& l_{\rm Pl} \leq r \leq l_{\rm dS} \, ,
\nonumber\\
&&l_{\rm Pl} \leq (\Delta x_{\rm canon.})_{\rm min} \leq \sqrt{l_{\rm Pl}l_{\rm dS}} \, ,
\nonumber\\
&&l_{\rm dS} \geq \Delta x_{\rm grav} \geq l_{\rm Pl} \, ,
\end{eqnarray}
respectively. 

These mass-scales also have interesting gravitational properties. 
To within numerical factors of order unity, the smallest possible mass $m_{\Lambda}$ is the the unique mass-scale satisfying the equation
\begin{eqnarray} \label{}
&& \lambda_{\rm C} \simeq r_{\rm grav}(\tau_0) \simeq (l_{\rm dS}^2r_{\rm S})^{1/3}  
\nonumber\\
&\iff& m \simeq m_{\Lambda} \simeq \sqrt{m_{\rm Pl}m_{\rm dS}} \, .
\end{eqnarray}
In other words, it is the only rest-mass whose quantum mechanical (Compton) radius is equal to its classical gravitational (turn-around) radius in the presence of dark energy. 
This gives a neat interpretation of the stability condition $m \geq m_{\Lambda}$ since, for smaller masses, the gravitational turn-around radius lies within the Compton wavelength of the particle. 
Considering the ranges of $r$ for which the canonical quantum uncertainty is greater or less than the gravitational uncertainty in the DE-UP-2 (\ref{DE-UP-2}), we have 
\begin{eqnarray} \label{}
&& (\Delta x_{\rm canon.})_{\rm min} \gtrless \Delta x_{\rm grav}
\nonumber\\
&\iff& r \gtrless r_{\rm eq} \equiv  2^{1/6}(l_{\rm dS}^2l_{\rm Pl})^{1/3}\frac{m}{m_{\rm Pl}} \, .
\end{eqnarray}
Setting $r_{\rm eq}$ equal to the present day turn-around radius of an object of mass $m$, and again neglecting numerical factors of order unity, then yields
\begin{eqnarray} \label{}
r_{\rm eq} &\simeq& r_{\rm grav}(\tau_0) \simeq (l_{\rm dS}^2r_{\rm S})^{1/3} 
\nonumber\\
\iff m &\simeq& m_{\rm Pl} \, .
\end{eqnarray}
For the critical mass $m_{\rm crit}$, we have
\begin{eqnarray} \label{}
r_{\rm eq} &\simeq& r_{\rm min} \simeq \lambda_{\rm C} 
\nonumber\\
&\simeq& (\Delta x_{\rm canon.})_{\rm min} \simeq \Delta x_{\rm grav} 
\nonumber\\
&\simeq& (\Delta x_{\rm total})_{\rm min}  \simeq (l_{\rm Pl}^2l_{\rm dS})^{1/3} \
\nonumber\\
\iff m &\simeq& m_{\rm crit} \simeq (m_{\rm Pl}^2m_{\rm dS})^{1/3} \, , 
\end{eqnarray}
where we recall that $r_{\rm min}$ is the probe distance that minimizes the total uncertainty, yielding Eq. (\ref{DE-UP-3}). 
The turn-around radius is then:
\begin{eqnarray} \label{}
r_{\rm grav}(\tau_0) \simeq (l_{\rm Pl}^4l_{\rm dS}^5)^{1/9} \, . 
\end{eqnarray}

In general, we note that, when $r$ is approximately equal to the present day turn-around radius, we have 
\begin{eqnarray} \label{}
(\Delta x_{\rm canon.})_{\rm min} &\simeq& (l_{\rm Pl}l_{\rm dS}\lambda_{\rm C})^{1/3}
\nonumber\\
\Delta x_{\rm grav} &\simeq& (l_{\rm dS}r_{\rm S}^2)^{1/3}
\nonumber\\
\iff r &\simeq& r_{\rm grav}(\tau_0) \, . 
\end{eqnarray}
Beyond this range, the {\it classical} gravitational influence of the particle is effectively negligible, in comparison to the repulsive effect of dark energy. 
In terms of space-time curvature, for $r \gtrsim r_{\rm grav}(\tau_0)$, the additional contribution to the total curvature due to $m$ is less than the background value $\sim \Lambda$. 
However, in order for the {\it quantum} gravitational influence of the particle to be considered negligible, it must induce metric fluctuations smaller than the background average, which are believed to be of order $\sim l_{\rm Pl}$ \cite{Garay:1994en}. 
We now consider this scenario in detail. 

To begin with, we note that, for $(\Delta x_{\rm canon.})_{\rm min}$ to be super-Planckian at $ r_{\rm grav}(\tau_0)$ requires $m \lesssim m_{\rm Pl}^2/m_{\rm dS}$, which is clearly satisfied for any physically realizable mass, up to and including the present day mass of the Universe. 
However, for $\Delta x_{\rm grav}$ to be super-Planckian at the turn-around radius requires $m \gtrsim m_{\Lambda} \simeq \sqrt{m_{\rm Pl}m_{\rm dS}}$. 
This result implies that metric fluctuations of order $\sim l_{\rm Pl}$ are associated with pure (empty) de Sitter space, since $m_{\Lambda}$ may also be interpreted as the mass of an effective dark energy particle \cite{Burikham:2015nma}. 
It therefore follows that, for any mass larger than $m_{\Lambda} $, the quantum gravitational influence of the particle at its turn-around radius will be {\it non-negligible}, in comparison to the magnitude of the background metric fluctuations, even if its classical gravitational influence may be ignored. 
{\it This is an important point, which may be relevant to future experimental attempts to distinguish between classical and quantum gravitational phenomenology predicted by the DE-UP model.}

Finally, it is straightforward to determine the ranges of $\Delta v$ (or equivalently $\Delta p$), $r$ and $m$ for which the three terms in the DE-UP-1 (\ref{DE-UP-1}) satisfy $\Delta x \geq \Delta x_{\rm recoil} \geq \Delta x_{\rm grav}$, or any other ordering. 
The results are summarized, for general values of $\alpha'$ and $\beta'$, in Table 1.

\begin{widetext}
\begin{center}
\begin{tabular}{|c|c|c|c|}
\hline
$ {\rm No.}$ & $r$ & $\Delta p$ & {\rm Order} \\
\hline
1 & $r \leq (2\beta'^2/\alpha')^{1/3}(l_{\rm Pl}l_{\rm dS}^2)^{1/3}(m/m_{\rm Pl})$ & $\Delta p \leq \frac{1}{2\beta'}\frac{\lambda_{\rm C}r}{l_{\rm Pl}l_{\rm dS}}m_{\rm Pl}c$ & $\Delta x_{\rm recoil} \leq \Delta x_{\rm grav} \leq \Delta x$ \\
\hline
2 & $``$ & $\frac{1}{2\beta'}\frac{\lambda_{\rm C}r}{l_{\rm Pl}l_{\rm dS}}m_{\rm Pl}c \leq \Delta p \leq \frac{1}{\sqrt{2\alpha'}}\frac{\hbar}{\sqrt{\lambda_{\rm C}r}}$ & $\Delta x_{\rm recoil} \leq \Delta x \leq \Delta x_{\rm grav}$  \\
\hline
3 & $``$ & $\frac{1}{\sqrt{2\alpha'}}\frac{\hbar}{\sqrt{\lambda_{\rm C}r}} \leq \Delta p \leq \frac{\beta'}{\alpha'}\frac{l_{\rm dS}l_{\rm Pl}^3}{\lambda_{\rm C}^2r^2}m_{\rm Pl}c$ & $\Delta x \leq \Delta x_{\rm recoil} \leq \Delta x_{\rm grav}$  \\
\hline
4 & $``$ & $\frac{\beta'}{\alpha'}\frac{l_{\rm dS}l_{\rm Pl}^3}{\lambda_{\rm C}^2r^2}m_{\rm Pl}c \leq \Delta p$ & $\Delta x \leq \Delta x_{\rm grav} \leq \Delta x_{\rm recoil}$  \\
\hline
5 & $r \geq (2\beta'^2/\alpha')^{1/3}(l_{\rm Pl}l_{\rm dS}^2)^{1/3}(m/m_{\rm Pl})$ &  $\Delta p \leq \frac{\beta'}{\alpha'}\frac{l_{\rm dS}l_{\rm Pl}^3}{\lambda_{\rm C}^2r^2}m_{\rm Pl}c$ & $\Delta x_{\rm recoil} \leq \Delta x_{\rm grav} \leq \Delta x$  \\
\hline
6 & $``$ & $\frac{\beta'}{\alpha'}\frac{l_{\rm dS}l_{\rm Pl}^3}{\lambda_{\rm C}^2r^2}m_{\rm Pl}c \leq \Delta p \leq \frac{1}{\sqrt{2\alpha'}}\frac{\hbar}{\sqrt{\lambda_{\rm C}r}}$ & $\Delta x_{\rm grav} \leq \Delta x_{\rm recoil} \leq \Delta x$  \\
\hline
7 & $``$ & $\frac{1}{\sqrt{2\alpha'}}\frac{\hbar}{\sqrt{\lambda_{\rm C}r}} \leq \Delta p \leq \frac{1}{2\beta'}\frac{\lambda_{\rm C}r}{l_{\rm Pl}l_{\rm dS}}m_{\rm Pl}c$  & $\Delta x_{\rm grav} \leq \Delta x \leq \Delta x_{\rm recoil}$  \\
\hline
8 & $``$ & $\frac{1}{2\beta'}\frac{\lambda_{\rm C}r}{l_{\rm Pl}l_{\rm dS}}m_{\rm Pl}c \leq \Delta p$ & $\Delta x \leq \Delta x_{\rm grav} \leq \Delta x_{\rm recoil}$   \\
\hline
\end{tabular}
\end{center}
\end{widetext}

For $r \lesssim (l_{\rm Pl}l_{\rm dS}^2)^{1/3}(m/m_{\rm Pl})$, low-momentum states are given by 1, intermediate-momentum states by 2-3 and high-momentum states by 4. 
As $r \rightarrow 0$, the limits in 1 and 4 tend to zero and infinity, respectively. 
For $r \gtrsim (l_{\rm Pl}l_{\rm dS}^2)^{1/3}(m/m_{\rm Pl})$, low-momentum states are given by 5, intermediate-momentum states by 6-7 and high-momentum states by 8. As $r \rightarrow \infty$, the limits in 5 and 8 tend to zero and infinity, respectively. 

Hence, for $r \lesssim (l_{\rm Pl}l_{\rm dS}^2)^{1/3}(m/m_{\rm Pl})$, $\Delta x_{\rm grav}$ may dominate $\Delta x_{\rm recoil}$, but not $\Delta x$, in the low-momentum regime, or $\Delta x$, but not $\Delta x_{\rm recoil}$, in the high-momentum regime. 
However, it may also dominate {\it both} in the intermediate-momentum regime. For $r \gtrsim (l_{\rm Pl}l_{\rm dS}^2)^{1/3}(m/m_{\rm Pl})$, the situation is similar in the `low-' and `high-' momentum regimes -- though these now correspond to different physical ranges of momentum uncertainty -- but is reversed in the intermediate regime, where $\Delta x_{\rm grav}$ is subdominant to both $\Delta x$ and $\Delta x_{\rm recoil}$.

From the point of view of future experiments, the $r \gtrsim (l_{\rm Pl}l_{\rm dS}^2)^{1/3}(m/m_{\rm Pl})$ regime is more accessible, and we are free to choose the ratio of the probe distance to the mass of the probe particle, $r/m$, to lie in this range. In this case, the very high- and very low-momentum regimes are where we may hope to observe modifications of canonical quantum dynamics. 
Nonetheless, the observability of these effects depends, ultimately, on the ratio of $\Delta x_{\rm grav}$, to the remaining (non-negligible) canonical uncertainty term.

When the DE-UP-1 (\ref{DE-UP-1}) is minimized with respect to $\Delta v$, yielding the DE-UP-2 (\ref{DE-UP-2}), $\Delta x \simeq \Delta x_{\rm recoil} \simeq (\Delta x_{\rm canon.})_{\rm min}$ and the value of $\Delta p$ is fixed in terms of $r$ by Eq. (\ref{DE-UP-1-mins}). 
Under these conditions, Table 1 reduces to the simple conditions
\begin{eqnarray} \label{Table-1_reduction}
r &\lesssim& (l_{\rm Pl}l_{\rm dS}^2)^{1/3}(m/m_{\rm Pl}) \iff (\Delta x_{\rm canon.})_{\rm min} \lesssim \Delta x_{\rm grav} \, , 
\nonumber\\
r &\gtrsim& (l_{\rm Pl}l_{\rm dS}^2)^{1/3}(m/m_{\rm Pl}) \iff (\Delta x_{\rm canon.})_{\rm min} \gtrsim \Delta x_{\rm grav} \, .
\nonumber\\
\end{eqnarray}
Hence, for $r \gtrsim (l_{\rm Pl}l_{\rm dS}^2)^{1/3}(m/m_{\rm Pl})$, $\Delta x_{\rm grav}$ is always sub-dominant to $ (\Delta x_{\rm canon.})_{\rm min}$. 
That said, the two need not, necessarily, be of comparable magnitude in order for $\Delta x_{\rm grav}$ to be detectable. 
The possibility of experimentally testing the DE-UP-1 (\ref{DE-UP-1}) using current technology will be addressed in a future publication.

Before concluding this subsection, we note that the minimum mass-scale $m_{\Lambda} = 4.832 \times 10^{-36}$g (\ref{m_Lambda}) is compatible with the current upper bound on the average neutrino mass obtained from the Planck mission data is $\langle m_{\nu}\rangle \leq 0.23$ eV $=4.100 \times 10^{-34}$ g \cite{PlanckCollaboration}. 
According to the arguments presented here, $m_{\Lambda}$ may be interpreted as the mass of the electron neutrino, which corresponds to the mass of the lightest {\it possible} neutral particle in a dark energy Universe with $\Lambda \simeq 10^{-56} {\rm cm^{-2}}$. 
An alternative interpretation of $m_{\Lambda}$ as the mass of a dark energy particle is discussed in Sec. \ref{sect4.2}.

\subsubsection{Application to charged particles} \label{sect3.3.2}

Let us now consider the implications of the DE-UP 
for charged particles. 
As shown in Sec. \ref{sect3.3}, combining the existence of a classical minimum density, which follows from the generalised Buchdahl inequalities for uncharged particles in the presence of dark energy \cite{Buchdahl:1959zz,Mak:2001gg,Boehmer:2005sm,Boehmer:2006fd}, with the standard expression for the Compton wavelength, gives rise to a minimum mass for compact, stable, gravitating, charge-neutral and quantum mechanical objects. 
Furthermore, this mass-scale is physically interesting as it is comparable to present day bounds on the mass of the lightest known particle, the electron neutrino \cite{Burikham:2015nma}. 
Combining the minimum-mass bound for neutral particles with the DE-UP also yields interesting results, since it implies that both the canonical and gravitational uncertainty terms, $(\Delta x_{\rm canon.})_{\rm min}$ and $\Delta x_{\rm grav}$, as well as the probe distance $r$, always remain super-Planckian.

Similarly, generalised Buchdahl inequalities exist for charged particles, both in the presence and absence of dark energy \cite{Burikham:2015sro,Boehmer:2007gq,Bekenstein:1971ej}. 
However, in this case, they fix only the minimum value of the radius-to-mass ratio, $R/m$, of a stable compact object, where $R$ is the {\it classical} radius.  
This bound may again be combined with the existence of a minimum quantum mechanical radius, $\lambda_{\rm C} = \hbar/(mc)$, and with the existence of a minimum total uncertainty given by Eq. (\ref{DE-UP-3}). 
The latter implies that the mass of the object may be written in terms of the critical mass, $m_{\rm crit} \simeq (m_{\rm Pl}^2m_{\rm dS})^{1/3}$, multiplied by an arbitrary constant $\alpha_{\rm Q}$, as in Eq. (\ref{critical_mass_rescale}). 

By combining all three mass bounds -- that is, by assuming that a charged particle exists in nature whose total uncertainty minimizes the DE-UP, according to Eq. (\ref{DE-UP-3}), whose classical radius satisfies the appropriate generalised Buchdahl inequalities \cite{Burikham:2015sro,Boehmer:2007gq,Bekenstein:1971ej}, and whose Compton radius is given by the canonical formula -- we fix the value of the free parameter $\alpha_{\rm Q}$ in terms of the the physical charge ($Q$) of the system. 
This, in turn, allows us to obtain an explicit expression for the mass $m$ in terms of $Q$ and the physical constants $\left\{G,c,\hbar,\Lambda\right\}$. 
Setting $Q=\pm e$ and evaluating this expression numerically, the mass-scale obtained is comparable to the measured value of the lightest charged particle, the electron \cite{Burikham:2015sro}. 
According to our procedure, this may be interpreted as the minimum {\it possible} mass for a compact, stable, gravitating, {\it charged} and quantum mechanical object, which also obeys the DE-UP proposed in Sec. \ref{sect3.2}. 

We proceed as follows. The generalized Buchdahl inequality for a charged compact object in the presence of a positive cosmological constant is \cite{Boehmer:2007gq}
\begin{eqnarray}  \label{GBI_Lambda_Q}
\frac{2Gm}{c^2 R}\geq \frac{3}{2}\frac{GQ^{2}}{c^4R^{2}} \left[\frac{1+\frac{1}{9} \frac{c^4\Lambda R^{4}}{GQ^{2}} 
- \frac{1}{54}\Lambda R^{2} + \frac{GQ^2}{18c^4R^2}} {1+\frac{GQ^{2}}{12c^4R^{2}}}\right].
\end{eqnarray}
Hence, for $R^2\Lambda \ll 1$, the effect of dark energy is subdominant to electrostatic repulsion and Eq. (\ref{GBI_Lambda_Q}) reduces to
\begin{eqnarray}  \label{GBI_Q}
\frac{2Gm}{c^2 R}\geq \frac{3}{2}\frac{GQ^{2}}{c^4 R^{2}} \left[\frac{1 + \frac{GQ^2}{18c^4R^2}}{1+ \frac{GQ^{2}}{12c^4R^{2}}}\right] \, .
\end{eqnarray}
This expression can be Taylor expanded to give
\begin{eqnarray}  \label{GBI_Q_Taylor}
\frac{2Gm}{c^2R}\geq \frac{3}{2}\frac{GQ^{2}}{c^4R^{2}} \Bigl(1-\frac{GQ^2}{36c^4R^2} + O(Q^2/R^2)^4 \Bigr) \, ,
\end{eqnarray}
so that, to leading order,  
\begin{eqnarray}  \label{R_class}
R \geq \frac{3}{4}\frac{Q^2}{mc^2} \, .
\end{eqnarray}
In this limit, to within numerical factors of order unity, we recover the standard expression for the classical radius of a `particle' with mass $m$ and charge $Q$, i.e. the radius at which the electrostatic potential energy associated with the object is equal to its rest energy, $mc^2$. 
In special relativity, this is roughly the radius the object would have \emph{if} its mass were due only to electrostatic potential energy. 
Nevertheless, Eq. (\ref{R_class}), which was originally obtained in \cite{Bekenstein:1971ej}, is a fully general-relativistic result. The fact the the standard formula for the classical radius of a charged particle is recovered via the Taylor expansion (\ref{GBI_Q_Taylor}) simply reflects the fact that Eqs. (\ref{GBI_Lambda_Q})-(\ref{GBI_Q}) remain valid, even in the weak gravity limit. 

Next, we note that a natural way to define the quantum gravitational regime for a fundamental particle is to require its positional uncertainty, due to combined canonical and quantum gravitational effects, to be greater than or equal to its classical radius, $\Delta x_{\rm total} = (\Delta x_{\rm canon.})_{\rm min} + \Delta x_{\rm grav} \geq R$. 
This is essentially the inverse of the requirement for classicality, that the macroscopic radius of an object be larger than its total positional uncertainty. 
Thus, the conditions
\begin{eqnarray}  \label{}
\lambda_{\rm C} \geq \Delta x_{\rm total} \geq R
\end{eqnarray}
correspond to a regime in which the particle behaves `quantum-gravitationally', but in which specific quantum gravitational effects are subdominant to the standard Compton uncertainty. 

Assuming that the total uncertainty takes its minimum possible value, given by the DE-UP-3 (\ref{DE-UP-3}), we may then set
\begin{eqnarray} \label{Delta_x_total-1}
(\Delta x_{\rm total})_{\rm min} = 3\left(\frac{\alpha'\beta'}{2}\right)^{1/3}(l_{\rm Pl}^2l_{\rm dS})^{1/3} = \gamma \lambda_{\rm C} \, ,
\end{eqnarray}
where $\gamma \leq 1$, in this regime. Likewise, we may set
\begin{eqnarray}  \label{}
\xi R = (\Delta x_{\rm total})_{\rm min} = 3\left(\frac{\alpha'\beta'}{2}\right)^{1/3}(l_{\rm Pl}^2l_{\rm dS})^{1/3} \, ,
\end{eqnarray}
where $\xi \geq 1$, if we expect the object to display no classical behaviour. Clearly,
\begin{eqnarray}  \label{gamma<xi}
\gamma \leq \xi \, ,
\end{eqnarray}
with equality holding if and only if $\gamma = \xi = 1$.

For convenience, we now rewrite the three independent expressions we have obtained for $m$ throughout the preceding sections of this work, namely
\begin{subequations}
\begin{align}
m &= \alpha_{\rm Q}m_{\rm crit.} = \left(\frac{1}{4\alpha'\beta'}\right)^{1/3} \alpha_{\rm Q}(m_{\rm Pl}^2m_{\rm dS})^{1/3} \, ,  \label{M-1A} \\
m &= \frac{\gamma}{3}\left(\frac{2}{\alpha'\beta'}\right)^{1/3} (m_{\rm Pl}^2m_{\rm dS})^{1/3} \, ,   \label{M-1B} \\
m &= \frac{\xi}{4}\left(\frac{2}{\alpha'\beta'}\right)^{1/3} \frac{Q^2}{q_{\rm Pl}^2} (m_{\rm Pl}^2m_{\rm dS})^{1/3} \, ,  \label{M-1C}
\end{align}
where $q_{\rm Pl} = \sqrt{\hbar c}$ is the Planck charge. 
Eqs. (\ref{M-1A}) and (\ref{M-1B}) are simply Eqs. (\ref{critical_mass_rescale}) and (\ref{Delta_x_total-1}) restated. 
Eq. (\ref{M-1C}) corresponds to saturating the bound in Eq. (\ref{R_class}) by assuming that $R = (\Delta x_{\rm total})_{\rm min}/\xi$ represents the value of the classical radius that minimizes the ratio $R/m$, for a sphere of mass $m$ and charge $Q$. 
(For the sake of generality, we have retained the 3/4 numerical factor from Eq.~(\ref{R_class}) but kept the numerical constants $\alpha'$ and $\beta'$ as unfixed parameters for now.)

Thus, $m$ in Eqs. (\ref{M-1A})-(\ref{M-1B}) is the mass of the body for which the total uncertainty of the object, given by the DE-UP, is minimized for $(\Delta v)_{\rm max} = \alpha_{\rm Q}^{-1}c$, whereas the $m$ in Eq. (\ref{M-1C}) is the mass of a body for which the classical bound (\ref{R_class}) is saturated. 
As shown in \cite{Burikham:2015sro}, this is also the radius at which the classical gravitational energy is minimized. 
We proceed by assuming the equivalence of the two masses, which is equivalent to assuming that the particle saturates {\it all} available bounds simultaneously. 

The resulting model has much in common with Dirac's extensive model of the electron \cite{Dirac:1962iy}, which was intended to remove singularities from the electric and gravitational fields of charged particles, except that, here, the classical electron is considered as a three-dimensional fluid sphere, rather than a two-dimensional shell. 
Nonetheless, the relevant Buchdahl bounds can be re-formulated in terms of two-dimensional (surface) quantities \cite{Burikham:2015sro}. 

By equating the three expressions for $m$ in Eqs. (\ref{M-1A})-(\ref{M-1C}), we may fix the relations between the three unknowns $\gamma$, $\xi$ and $\alpha_{\rm Q}$, explicitly. 
For our purposes, the key point is that, for $\xi \sim \mathcal{O}(1)$ (i.e. when $(\Delta x_{\rm total})_{\rm min} \simeq R$, its minimum possible value), we have $\gamma \simeq \alpha_{\rm Q} \simeq Q^2/q_{\rm Pl}^2$. 

Equations (\ref{M-1B}) and (\ref{M-1C}) immediately imply
\end{subequations}
\begin{eqnarray}  \label{gamma}
\frac{Q^2}{q_{\rm Pl}^2} = \frac{4}{3} \frac{\gamma}{\xi} \lesssim 1 \, ,
\end{eqnarray}
or, equivalently,
\begin{eqnarray}  \label{Q<q_P}
Q \lesssim q_{\rm Pl} \, .
\end{eqnarray}
This gives an interesting and self-consistent interpretation of the Planck charge, $q_{\rm Pl}$, as the maximum possible charge of a stable, gravitating, quantum mechanical object, obeying the DE-UP. 
The bound (\ref{Q<q_P}) may also be obtained in a more direct way by combining the general relativistic result (\ref{R_class}) with canonical quantum theory. Rewriting this as $Q^2 \leq (4/3)q_{\rm Pl}^2 Rmc^2$ and taking the limit $R \rightarrow \lambda_{\rm C}$ yields the same result.

For the sake of concreteness, we now set
\begin{eqnarray} \label{}
\gamma = \frac{3}{2}\frac{Q^2}{q_{\rm Pl}^2}  \, , \quad \xi = 2 \, , 
\end{eqnarray}
yielding
\begin{eqnarray}  \label{alpha_Q}
\alpha_{\rm Q} = \frac{Q^2}{q_{\rm Pl}^2} \, ,
\end{eqnarray}
and choose the values $\alpha' = 1/2\sqrt{2}$ and $\beta'  = \sqrt{2}$ obtained previously (\ref{alpha'beta'}), so that 
\begin{eqnarray}  \label{}
m = \alpha_{\rm Q}m_{\rm crit} = 2^{-1/3}\alpha_{\rm Q}(m_{\rm Pl}^2m_{\rm dS})^{1/3}  \, ,
\end{eqnarray}
and 
\begin{eqnarray}  \label{}
r_{\rm min} = (4\sqrt{2})^{1/3}\alpha_{\rm Q}(l_{\rm Pl}^2l_{\rm dS})^{1/3}  \, .
\end{eqnarray}
We then have 
\begin{eqnarray}  \label{}
R &=& \frac{1}{2}(\Delta x_{\rm total})_{\rm min} 
\nonumber\\
&=& \left(\frac{27}{4}\right)^{1/3}\alpha_{\rm Q}(l_{\rm Pl}^2l_{\rm dS})^{1/3}  = \frac{3}{4}\alpha_{\rm Q}\lambda_{\rm C} \, .
\end{eqnarray}
Though the precise numerical factors chosen here are to some degree arbitrary, we see that, for $\xi \sim \mathcal{O}(1)$, the following order of magnitude relations hold:
\begin{eqnarray}  \label{oom-relns-1}
m \simeq \alpha_{\rm Q}(m_{\rm Pl}^2m_{\rm dS})^{1/3} \, , \quad r_{\rm min} \simeq \alpha_{\rm Q}^2 \lambda_{\rm C} \, ,
\end{eqnarray}
\begin{eqnarray}  \label{oom-relns-2}
(\Delta x_{\rm total})_{\rm min} \simeq R \simeq \alpha_{\rm Q} \lambda_{\rm C} \, ,
\end{eqnarray}
where $\alpha_{\rm Q}$ is given by Eq. (\ref{alpha_Q}).

Keeping in mind the generalized measurement scheme outlined in Sec. \ref{sect3.2}, the physical picture we obtain is as follows. 
A particle of mass $m$ and charge $Q$ `measures' the distance to its outermost horizon, the de Sitter radius, by means of a two-stage photon exchange. 
In the first stage, photons (either real or virtual) are exchanged between the particle CoM and a `detector' at $r$. 
The `detector' simultaneously (or near simultaneously, within $\sim r/c$) receives real or virtual photons from the de Sitter horizon. 
{\it The minimum total uncertainty in the position of the particle is also the minimum uncertainty in the measurement of $l_{\rm dS}$.} 

However, as discussed in Sec. \ref{sect3.2}, and at length in the Appendix, for $r < \lambda_{\rm C}$ the `detector' is simply the particle itself and the first part of the `measurement' corresponds to a self-interaction. 
What the relations above show is that the total uncertainty given by the DE-UP-1 (\ref{DE-UP-1}) obtains its minimum possible value, given by the DE-UP-3 (\ref{DE-UP-3}), when the charge-squared to mass ratio of the particle $Q^2/m$, and the corresponding self-interaction distance $r_{\rm min}$, are fixed according to Eq. (\ref{oom-relns-1}). 
Under these circumstances, the order of magnitude values of $R \simeq Q^2/(mc^2)$, $(\Delta x_{\rm total})_{\rm min}$ and $\lambda_{\rm C}$ are also fixed, yielding a strict hierarchy of length-scales associated with $m$. 
These are related via the parameter $\alpha_{\rm Q} = Q^2/q_{\rm Pl}^2$ according to Eq. (\ref{oom-relns-2}). 

That the minimum uncertainty in the position of the particle is larger than the probe distance $r$ need not concern us, since $r_{\rm min}$ may be associated with the energy scale of the self-interaction via the usual Compton formula, giving $E_{\rm max} \simeq \hbar c/r_{\rm min} \simeq (q_{\rm Pl}^4/Q^2)/\lambda_{\rm C}$ as a natural UV `cut-off' in the DE-UP model. 
Though not strictly a cut-off, attempting to probe self-gravitating particles on scales $r < r_{\rm min}$ ($E > E_{\rm max}$) is self-defeating, since this only increases $\Delta x_{\rm total}$.

In this picture, a charged particle that interacts with its environment (including self-interactions) over the range $r_{\rm min} \leq r \leq l_{\rm dS}$ must possess a charge-squared to mass ratio that satisfies the bound:
\begin{eqnarray}  \label{BOUND}
\frac{Q^2}{m} &\lesssim& \left(\frac{3\hbar^2G^2c^6}{\Lambda}\right)^{1/6} = 3.147 \times 10^8 \, {\rm Fr^2g^{-1}} 
\nonumber\\
&\simeq& \frac{e^2}{m_e} = 2.533 \times 10^{8} \, {\rm Fr^2g^{-1}} \, .
\end{eqnarray}
This is obtained simply by rewriting the expression for $m$ Eq. (\ref{oom-relns-1}) and reinserting the directional inequality originally present in Eq. (\ref{R_class}). 
Thus, it is straightforward to see that, to within numerical factors of order unity,  saturating the bound (\ref{BOUND}) is equivalent to setting $Q^2 = e^2$, which yields the correct order of magnitude value of the electron mass, i.e.
\begin{eqnarray}  \label{}
&& m = \alpha_{e}(m_{\rm Pl}^2m_{\rm dS})^{1/3} = 7.332 \times 10^{-28} \, {\rm g}
\nonumber\\
&\simeq& m_{e} = 9.109 \times 10^{-28} \, {\rm g} \, ,
\end{eqnarray}
where $\alpha_{e} = e^2/q_{\rm Pl}^2$ is the usual fine structure constant. 

Alternatively, Eq. (\ref{BOUND}) may be rewritten to obtain an upper bound on the value of $\Lambda$, using the measured value of $m_{e}$, giving 
\begin{eqnarray}  \label{Lambda}
\Lambda \lesssim \frac{m_{e}^6G^2}{\alpha_{e}^6\hbar^4} \simeq 1.366 \times 10^{-56} {\rm cm^{-2}} \, ,
\end{eqnarray}
which is close to the best-fit value obtained from current cosmological observations \cite{Betoule:2014frx,PlanckCollaboration}. 
The result (\ref{Lambda}) was previously obtained by Harko and Boehmer in \cite{Boehmer:2006fd}, in which it was expressed in the form $\Lambda \simeq l_{\rm Pl}^4/r_{e}^6$, where $r_{e} = e^2/(m_{e}c^2)$ is the classical electron radius, and justified on the basis of a `Small Number Hypothesis' (SNH), by analogy with Dirac's Large Number Hypothesis (LNH), which posits that the {\it numerical equality between two very large quantities with a very similar physical meaning cannot be a simple coincidence} \cite{Dirac1937,Dirac1979}.

We stress, however, that in this work, the identification (\ref{Lambda}) is \emph{not} based on numerical coincidence. 
Rather, our requirement that the total uncertainty $\Delta x_{\rm total}$, incorporating canonical quantum and gravitational effects according to the DE-UP, be minimized for a stable, compact, charged, gravitating and quantum mechanical object, realised in nature, leads inevitably to Eq. (\ref{Lambda}). 

\section{Cosmological consequences of the DE-UP} \label{sect4}

\subsection{Holography} \label{sect4.1}

It is straightforward to see that, for any particle which minimizes the total uncertainty given by the DE-UP according to Eq. (\ref{DE-UP-3}), a holographic relation holds between the bulk and the boundary of the Universe. 
Specifically, 
\begin{eqnarray}  \label{holog}
\left(\frac{(\Delta x_{\rm total})_{\rm min}}{l_{\rm dS}}\right)^3 = \frac{l_{\rm Pl}^2}{l_{\rm dS}^2} = N = 1.030 \times 10^{122} \, , 
\end{eqnarray}
so that the number of Planck sized `bits' on the de Sitter boundary is equal to the number of minimum-volume cells, $V_{\rm cell} \simeq (\Delta x_{\rm total})_{\rm min}^3$, in the bulk \cite{Burikham:2015sro}.  

\subsection{Dark energy particles?} \label{sect4.2}

As shown in \cite{Burikham:2015nma}, $m_{\Lambda}$ may also be interpreted as the effective mass of a dark energy particle. In this picture, the dark energy field is composed of a `sea' of quantum particles, each occupying a volume $\sim l_{\Lambda}^3$. 
Under these conditions, and if dark energy particles are charge-neutral but {\it fermionic}, the usual laws of quantum mechanics imply that they will readily pair-produce. 
However, this is {\it impossible} without a concomitant expansion in space itself. (In short, `empty' space is, in fact, {\it full} of dark energy particles.) 
Borrowing a term from basic chemistry to describe this state of affairs, we may say that the space is {\it saturated}. It is straightforward to see that, if the probability of pair-production remains constant, the scale factor of the Universe will grow exponentially since the number of particles produced in any given volume, per unit time, is proportional to the volume itself. 
This leads naturally to a de Sitter-type expansion, $da/d\tau \propto a$, in which the {\it macroscopic} dark energy energy density remains constant, in spite of spatial expansion. For particles of mass $m_{\Lambda}$, the additional (positive) energy of the newly created rest mass is exactly counterbalanced by the additional (negative) energy of its gravitational field, which may be seen by considering the Komar mass \cite{strong_grav}. 

However, if this picture is correct, we may expect `empty' three-dimensional space to exhibit {\it granularity} on scales $\sim l_{\Lambda}$. 
For this reason, it is particularly intriguing that recent experiments provide tentative hints of fluctuations in the strength of the gravitational field on scales comparable to $ l_{\Lambda}$, which is of order $\sim 0.1$ mm \cite{Perivolaropoulos:2016ucs,Antoniou:2017mhs}. 
Though many theoretical models may account for this, including those exhibiting spatial variation of the gravitational constant $G$, the influence of dark energy particles on sub-millimetre gravitational interactions cannot be discounted {\it a priori}. 

Furthermore, we note that, if the probability of a single cell of space `pair-producing' within a time interval $\Delta \tau = t_{\rm Pl} = l_{\rm Pl}/c$, due to the production of dark energy particles, is given by
\begin{eqnarray}  \label{}
&&P(\Delta V = +V_{\rm cell}|V_0=V_{\rm cell},\Delta \tau =t_{\rm Pl}) = N^{-1/2}
\nonumber\\ 
&& = \frac{V_{\rm Pl}}{V_{\rm cell}} = \frac{l_{\rm Pl}}{l_{\rm dS}} \simeq \left(\frac{\hbar G \Lambda}{3c^3}\right)^{1/2} \simeq 9.851 \times 10^{-62} \, ,
\end{eqnarray}
where $V_0$ denotes the initial volume, this leads naturally to a de Sitter-type expansion, modeled by the differential equation 
\begin{eqnarray} \label{dS_exp-1}
\frac{d a^3}{d\tau} \simeq \frac{N^{-1/2} a^3}{t_{\rm Pl}} = \frac{l_{\rm Pl}}{l_{\rm dS}}\frac{a^3}{t_{\rm Pl}} \, ,
\end{eqnarray}
or, equivalently \cite{strong_grav}:
\begin{eqnarray}\label{dS_exp-2}
\frac{d a}{d\tau} \simeq c\sqrt{\frac{\Lambda}{3}}a \, ,  \quad a(\tau) \simeq a_0e^{-c\sqrt{\Lambda/3}\tau} \, ,
\end{eqnarray}

The production of a single dark energy particle then requires the production of $n_{\rm cell} = V_{\Lambda}/V_{\rm cell} \simeq l_{\Lambda}^3/(l_{\rm Pl}^2l_{\rm ds}) = N^{1/4}$ cells of space which, in turn, implies that the probability of a dark energy particle pair-producing within $\Delta \tau = t_{\rm Pl}$ is given by
\begin{eqnarray}  \label{}
P(\Delta V &=& +V_{\rm \Lambda}|V_0=V_{\rm \Lambda},\Delta \tau = t_{\rm Pl}) \simeq N^{-3/4} 
\nonumber\\
&=& \left(\frac{l_{\rm ds}}{l_{\rm Pl}}\right)^{-3/2} \simeq 3.083 \times 10^{-92} \, .
\end{eqnarray}
Since there are $n_{\rm DE} \simeq l_{\rm dS}^3/l_{\Lambda}^3 = N^{3/4}$ dark energy particles within the de Sitter horizon, this implies that one dark energy particle is produced {\it somewhere} in the observable Universe during every Planck-time interval. Remarkably, this rate of pair-production is capable of giving rise to the accelerated expansion of the Universe observed at the current epoch. 

In this model, the observed vacuum energy is really the energy associated with the dark energy field: its fundamental dynamics remain unknown, but are assumed to be associated with the mass-scale $m_{\Lambda}$, and excitations of the vacuum state correspond to the production of charge-neutral particles with this mass. 
Thus, $\lambda_{\rm C}(m_{\Lambda}) = l_{\Lambda}$ provides a natural a cut-off for the field modes -- with higher-energy excitations yielding pair-production of dark energy particles throughout space -- so that
\begin{eqnarray} \label{rho_vac}
\rho_{\rm vac} \simeq \frac{\hbar}{c} \int_{1/l_{\rm dS}}^{1/l_{\Lambda}} \sqrt{k^2 + \left(\frac{2\pi}{l_{\Lambda}}\right)^2} d^3k 
\nonumber\\
\simeq \frac{m_{\rm Pl}l_{\rm Pl}}{l_{\Lambda}^{4}} \simeq \frac{\Lambda c^2}{G} \simeq 10^{30} \ {\rm gcm^{-3}} \, .
\end{eqnarray}
The precise dynamics, or `true' nature of the dark energy field, are essentially unobservable at the current epoch as the field remains `trapped' in a Hagedorn-type phase in which any increase in kinetic energy, {\it even that caused by random collisions between neighbouring dark energy particles due to quantum uncertainty}, results in pair-production rather than an increase in temperature / kinetic energy. 
(The interested reader is referred to \cite{strong_grav} for a more in-depth discussion of this point.)

The temperature associated with the field is therefore constant, on large scales, and is comparable to the present day temperature of the CMB,
\begin{eqnarray} \label{T_Lambda}
T_{\Lambda} \equiv \frac{m_{\Lambda}c^2}{8 \pi k_{\rm B}} \simeq 2.27 \ {\rm K} \simeq T_{\rm CMB} = 2.73 \ {\rm K} \, . 
\end{eqnarray}
Here, the factor of $8\pi$ is included by analogy with the expression for the Hawking temperature, 
\begin{eqnarray} \label{T_H}
T_{\rm H} \equiv \frac{c^2}{8 \pi k_{\rm B}}\frac{m_{\rm Pl}^2}{m} \, ,
\end{eqnarray}
so that $T_{\Lambda} \equiv T(m_{\Lambda}) = T_{\rm H}(m_{\Lambda}')$, where $m_{\Lambda}' = m_{\rm Pl}^2/m_{\Lambda}$ denotes the mass dual to $m_{\Lambda}$. 

Though this may seem like another miraculous coincidence, in the dark energy model implied by the DE-UP it is simply a restatement of the standard coincidence problem of cosmology, whereby the Universe begins a phase of accelerated expansion at the present epoch, when $r_{\rm U} \simeq l_{\rm dS}$ and $\Omega_{\rm M} \simeq \Omega_{\Lambda}$ and, hence, $T_{\rm CMB} \simeq T_{\Lambda}$. 
The coincidence remains: why do {\it we} live at precisely this epoch? However, no new coincidences are required, in order to explain Eq. (\ref{T_Lambda}) in the context of the DE-UP.

\subsection{Time-variation of `fundamental' constants?} \label{sect4.3}

At epochs prior to the present day, $\tau \lesssim \tau_0$, the cosmic horizon is smaller than the de Sitter radius and, strictly, we must substitute $l_{\rm dS} \rightarrow {\cal H}(\tau)/c$ in Eq. (\ref{DE-UP-1}) and all subsequent formulae derived from it. 
In this case, the upper bound on the charge-squared to mass ratio for stable charged particles obeying the DE-UP, Eq. (\ref{BOUND}), is {\it lowered} and drops {\it below} the charge-squared to mass ratio of present day electrons, yielding
\begin{eqnarray}  \label{BOUND*}
\frac{Q^2}{m}(\tau) \lesssim c^2(l_{\rm Pl}^2c/{\cal H}(\tau))^{1/3} \, . 
\end{eqnarray}
This corresponds to a minimum holographic cell radius 
\begin{eqnarray}  \label{BOUND**}
(\Delta x_{\rm total})_{\rm min}(\tau)  \simeq (l_{\rm Pl}^2c/{\cal H}(\tau))^{1/3} \, ,
\end{eqnarray}
which is identical to the MLUR for an expanding Universe recently suggested by Ng \cite{Ng:2016wzj}.
Similar arguments apply to the minimum mass for neutral particles, which is required to ensure that $(\Delta x_{\rm canon.})_{\rm min}$, $\Delta x_{\rm grav}$ and $r$ each remain super-Planckian, yielding
\begin{eqnarray}  \label{BOUND***}
m(\tau) \gtrsim \sqrt{m_{\rm Pl}m_{\mathcal{H}}(\tau)} \, ,
\end{eqnarray}
where $m_{\mathcal{H}}(\tau) = \hbar{\cal H}(\tau)/c^2$ is the Compton mass associated with the horizon distance at time $\tau$. 

However, identifying the epoch-dependent mass limits  (\ref{BOUND*})-(\ref{BOUND***}) with the the low-energy rest-masses of charged and neutral particles, respectively, is problematic, since this implies strong violation of Lorentz invariance, and, consequently, of energy and momentum conservation. 
Nonetheless, the resulting time-dependent quantities could be interpreted, not as the limiting low-energy values of $m_e$, $m_{\nu_e}$ and $e$, but as renormalized values. 
Since the standard model couplings and masses are energy-dependent due to renormalization group flow and, since changing ${\cal H}(\tau)$ changes both the UV and IR cut-offs for interactions in the DE-UP model, the relationship between these two energy-dependent factors may be non-trivial. 

What is clear is that, within the limits of the non-relativistic (i.e., non-Lorentz invariant) theory formulated here, such questions are very difficult to answer. 
To satisfactorily address them, we need to go beyond the non-relativistic approximation. 
Nonetheless, it is interesting to note the similarity of the minimum particle mass (\ref{BOUND***}) with the (running) dark energy mass-scale predicted by `agegraphic'  \cite{Cai:2007us,Wei:2007ty} and holographic \cite{Wei:2007ig} dark energy models previously proposed in the literature.

It is also interesting to note that the relation (\ref{Lambda}) was originally found by Nottale \cite{Nottale1993} using a renormalization group approach. 
He argued that, like other fundamental `constants', the cosmological constant is in fact a scale-dependent quantity, obeying an (as yet unknown) renormalization group equation. 
If so, its present day value may be split into a `bare' gravitational part plus a scale-dependent part, corresponding to the quantum mechanical vacuum energy, i.e. $\Lambda(r) = \Lambda_{\rm G} + \Lambda_{\rm QM}(r)$. 
Following Zel'dovich \cite{Zel'dovich:1968zz}, who noted that the bare zero-point energy is unobservable, he then argued that the observable contribution is given by the gravitational energy of virtual $e^{-}$-$e^{+}$ pairs, continually created and annihilated in the vacuum. 
Using only these assumptions, he obtained Eq. (\ref{Lambda}).

This is a remarkable achievement. 
However, we note that Nottale's argument implicitly assumed that the `gravitational cut-off', i.e., the UV cut-off $r_{\rm min}$ in the expression for the gravitational self-energy of a particle pair-produced in the vacuum, is equal to the average inter-particle distance. 
A priori, there is no reason why this should be the case. 
In fact, the most natural assumption for virtual particles pair-produced in the vacuum is that the average inter-particle distance is comparable to the Compton wavelength, in this case $\lambda_{\rm C}(m_{\rm e})$. 
In this scenario, the gravitational self-energy is
\bea \label{N-1**}
E_{\rm grav} \simeq \frac{G\hbar^2}{c^2r_{\rm min}^3} \, , 
\eea
but the vacuum density is
\bea \label{N-2*}
\rho_{\rm vac}(r) \simeq \rho_{\rm Pl}\frac{l_{\rm Pl}^6}{\lambda_{\rm C}^3(m_{\rm e})r_{\rm min}^3} \simeq \frac{m_{\rm e}^3c^2}{m_{\rm Pl}r_{\rm min}^3}\, ,
\eea
Interestingly, setting $r_{\rm min} \simeq \alpha_{\rm e}^2\lambda_{\rm C}(m_{\rm e})$, the minimum `probe' distance for a particle of charge $\pm e$ predicted by the DE-UP (\ref{oom-relns-1}), and identifying $\rho_{\rm vac} \equiv \rho_{\Lambda}$ in Eq. (\ref{N-2*}), also yields Eq. (\ref{Lambda}). 
In this sense, the predictions of the DE-UP model may also be considered as compatible with Nottale's analysis.
In an expanding Universe, a vacuum energy of the form $\rho_{\rm vac} \simeq 3{\cal H}^{2}(\tau)/(8\pi G)$, coupled with a Nottale-type analysis, also gives rise to Eqs. (\ref{BOUND*})-(\ref{BOUND***}). 
The interested reader is refered to \cite{Lake2017} for further details. 

Finally we note that, throughout this paper, we have considered the simplest dark energy scenario that is consistent with existing data sets, namely, dark energy in the form of a cosmological constant. 
However, dynamical dark energy models cannot be ruled out on the basis of current observations. 
While the cosmological constant model corresponds to a constant equation of state (EoS) parameter, $w_{\rm DE} = p_{\rm DE}/\rho_{\rm DE} = -1$, dynamical dark energy scenarios including quintessence ($w_{\rm DE} \geq -1$), phantom dark energy ($w_{\rm DE} \leq -1$) and quintom models, in which the dark energy EoS parameter evolves across the cosmological constant boundary, remain consistent with current data sets.
The latter may even be marginally favoured by existing data, see \cite{Cai:2009zp}.
 
Furthermore, while these four scenarios broadly classify the different types of dynamical dark energy theory, each may be realised by a whole host of specific models. 
Dynamical dark energy theories that have been extensively studied in the literature include $f(R)$ theories \cite{DeFelice:2010aj}, Lanczos-Lovelock gravity \cite{Padmanabhan:2013xyr}, general scalar-tensor (Horndeski) models \cite{Bellini:2015xja}, which include Galileon \cite{Barreira:2014jha} and chameleon models \cite{Burrage:2017qrf} as special cases, torsional models - including Einstein-Cartan \cite{Trautman:2006fp}, teleparallel and $f(T)$ gravity \cite{Cai:2015emx} -  massive gravity \cite{DeFelice:2017yde}, braneworlds \cite{Maartens:2010ar} and theories of varying `fundamental' constants \cite{Fritzsch:2016ewd}, though this list is by no means exhaustive, see \cite{Clifton:2011jh}.
  
Clearly, the realisation of any of these, or even more exotic models, in nature, should have profound implications for dark energy-induced MLURs, especially at times prior to the present epoch. 
Though it is beyond the scope of the present paper to investigate these effects in detail, the study of MLURs induced by dynamical dark energy models may be a fruitful avenue of future research.

\section{Conclusions} \label{sect5}

We have proposed a new minimum length uncertainty relation (MLUR), defined by Eqs. (\ref{DE-UP-1})-(\ref{DE-UP-3}), which incorporates both canonical quantum and gravitational effects in the presence of dark energy, given by a positive cosmological constant $\Lambda >0$. 
In this model $\Lambda$ is assumed to be a fundamental constant of nature, giving rise to a constant minimum energy density $\rho_{\Lambda} \propto \Lambda$ at {\it all} points in space. 
The new relation, termed the dark energy uncertainty principle, or DE-UP, is structurally similar to the MLUR proposed by K{\' a}rolyh{\' a}zy, Eq. (\ref{Karolyhazy_MLUR}) \cite{Karolyhazy:1966zz,KFL}, and reproduced by Ng and van Dam using alternative arguments, Eq. (\ref{MLUR-1}) \cite{Ng:1993jb,Ng:1994zk}. 

However, while both derivations of Eq. (\ref{Karolyhazy_MLUR})/(\ref{MLUR-1}) considered gravitational corrections to canonical (non-gravitational) quantum theory, each did so under the assumption that the background space-time was both asymptotically flat {\it and} static. 
Though these assumptions are valid in many physically interesting regimes, it is clear that the discovery of dark energy \cite{Reiss1998,Perlmutter1999} gives rise to a new fundamental length-scale in physics, namely, the de Sitter horizon $l_{\rm dS} \sim 1/\sqrt{\Lambda}$, as well as to an associated minimum curvature given by $\Lambda$. 
On cosmological time-scales, it is also clear that the effects of universal expansion on local physics must somehow be taken into account \cite{Giulini:2013zha}. 
In the DE-UP, the effects of minimum curvature and of a maximum horizon distance for {\it all} observers, including quantum mechanical particles, are explicitly accounted for, and the effects of universal expansion are incorporated into the MLUR. 

At a technical level, our derivation of the DE-UP closely resembles Ng and van Dam's derivation of Eq. (\ref{MLUR-1}). The primary difference is that, whilst they assumed the gravitational uncertainty of a fundamental particle is given by its Schwarzschild radius, we assume it is, instead, given by the irremovable quantum uncertainty inherent in a `measurement' of the particle's horizon distance, $\sim {\cal H}(\tau)/c$, where $\tau$ is the cosmic time. 
The physical basis for this assumption is straightforward. 
Since, {\it classically}, the distance between the particle and its horizon is exact, any quantum uncertainty inherent in the measurement of ${\cal H}(\tau)/c$ is equivalent to an irremovable uncertainty in the position of the particle itself.  

Hence, in order to estimate the uncertainty in a measurement of the horizon distance, induced by the effects of the particle's gravitational field, we assumed a simple relationship between the classical perturbation of the space-time line element, induced by the presence of the particle ($\Delta s_{\rm pert}$), and the quantum mechanical spread in the superposition of background geometries ($\Delta s$), i.e. $\Delta s_{\rm pert} \simeq \Delta s$ (\ref{ds_assumption}). (See Fig. 3.) 
This, in turn, allowed us to demonstrate the equivalence of K{\' a}rolyh{\' a}zy's procedure for `resolving' space-time intervals, using quantum mechanical particles as `probes', and the interaction of a particle with its outermost horizon. 
Whilst, clearly, this assumption cannot remain valid for macroscopic objects, and must break down at some critical mass- / length-scale, it leads to a number of interesting and physically viable predictions based on the DE-UP (\ref{DE-UP-1})-(\ref{DE-UP-3}). 
We note that the scale at which this assumption becomes invalid may be naturally related to K{\' a}rolyh{\' a}zy's concept of a coherence cell \cite{Karolyhazy:1966zz,KFL}, though a detailed investigation of the this possibility lies beyond the scope of this work.

Applying the DE-UP to neutral particles, and requiring {\it all} potentially observable length-scales to remain super-Planckian, implies the existence of minimum mass-scale in nature, which can be expressed in terms of the fundamental constants $\left\{G,c,\hbar,\Lambda\right\}$. Furthermore, this mass-scale can be derived independently by combining classical minimum mass bounds for stable compact objects, in the presence of dark energy, with the simple requirement of the existence of a Compton wavelength \cite{Burikham:2015sro}. The DE-UP is thus naturally consistent with known gravitational {\it and} quantum mechanical effects, as well as with the presumed minimum resolution due to quantum gravitational effects at the Planck scale. 

Evaluating the minimum mass for neutral particles numerically, it is of order $10^{-3}$ eV, and is consistent with current experimental bounds on the mass of the electron neutrino obtained from Planck satellite data \cite{PlanckCollaboration}. 
This mass-scale may also be interpreted as the mass of a dark energy particle \cite{Burikham:2015sro}. 
Such a model implies that, though the dark energy density is approximately constant on large scales, it may become {\it granular} on length-scales of order 0.1 mm, the associated Compton wavelength. 
With this in mind, it is particularly intriguing that recent submilimetre tests of Newtonian gravity reveal tentative evidence for periodic variation in the gravitational field strength over precisely this length-scale \cite{Perivolaropoulos:2016ucs,Antoniou:2017mhs}. 

Applying the DE-UP to electrically charged particles, we defined the quantum gravity regime as the regime in which the minimum total uncertainty, including both canonical quantum and gravitational contributions, is larger than the classical radius but smaller than the Compton radius. 
Evaluating this condition for a particle of charge $e$, at the current cosmological epoch $\tau_0$, we obtained the minimum mass of a stable, compact, charged, gravitating and quantum mechanical object, obeying the DE-UP, in terms of the constants $\left\{G,c,\hbar,\Lambda,e\right\}$. Numerically, this is of order $10^{-28}$ g, which is consistent with the current measured value of the electron mass $m_e$ \cite{Burikham:2015nma}.
At all epochs, the DE-UP implies the existence of a holographic relation between the bulk and the boundary of the Universe, in which the number of minimum-uncertainty `cells' in the bulk equals the number of Planck sized `bits' on the boundary.   

Although the DE-UP proposed herein suffers from a number of drawbacks, including an incomplete picture of the communication between a particle and its cosmological horizon, and a reliance on the assumption of an intimate connection between classical perturbations and space-time superpositions, we believe that it yields sufficiently interesting predictions to be worth further study. 
With future high-precision quantum experiments in mind, therefore, we have identified two regimes, listed in Table 1, in which the gravitational uncertainty term in the DE-UP dominates {\it at least one} of the two positional uncertainty terms obtained in canonical QM. 

\acknowledgments

Thanks to Wansuree Massagram for stimulating discussions, encouragement, and valuable artistic contributions to Fig. 2. 
This work is supported by Singapore Ministry of Education Academic Research Fund Tier 1 Project No. RG106/17.

\renewcommand{\theequation}{A-\arabic{equation}}
\setcounter{equation}{0} 
\section*{Appendix: Conceptual issues -- is $r < (\Delta x_{\rm total})_{\rm min}$ physical?} \label{Appendix}

In this Appendix, we argue that, while quantum effects on length-scales $r < (\Delta x_{\rm total})_{\rm min} < \lambda_{\rm C}$ cannot be treated {\it explicitly} within the non-relativistic theory, as these correspond to energies for which relativistic quantum effects become important, they are nonetheless physical. 
Likewise, we argue that, though our na{\" i}ve model predicts a minimum probe distance for charged particles, $r_{\rm min} \simeq \alpha_{\rm Q}^2\lambda_{\rm C} \simeq \alpha_{\rm Q}(\Delta x_{\rm total})_{\rm min}$ ($\alpha_{\rm Q} \leq 1$), at which relativistic quantum gravitational effects become important, this too may be regarded as physical, even if the full relativistic theory of quantum gravity required to treat it in detail is lacking.

Hence, if $\Delta x_{\rm total}(r)$ represents the total positional uncertainty of a quantum particle, as seen by an observer located at a distance $r$ from its CoM or, equivalently, the irremovable uncertainty in any measurement of $l_{\rm dS}$, obtained via the two-stage measurement process outlined in Sec. \ref{sect3.2}, we may ask the question: is it physically meaningful to consider $r < \Delta x_{\rm total}(r)$?
In general, for a particle of a given mass $m$, we may solve the inequality $r \lesssim \Delta x_{\rm total}(r)$ to find the critical value of $r$, below which this condition holds. 

Intriguingly, and at first sight somewhat bizarrely, the analysis presented in Secs. \ref{sect3}-\ref{sect4} suggests that $\Delta x_{\rm total}(r)$, given by Eq. (\ref{DE-UP-1}), is minimized for $r_{\rm min} \simeq \alpha_{\rm Q}(\Delta x_{\rm total})_{\rm min} \simeq \alpha_{\rm Q}(l_{\rm Pl}^2l_{\rm dS})^{1/3}$, where $\alpha_{\rm Q} = Q^2/q_{\rm Pl}^2 \leq 1$. 
In other words, when the uncertainty in the measured value of $\Delta x_{\rm total}(r)$ is as small as it can be, it is larger than the `probe' distance $r$. 
To interpret this result correctly, we must reconsider the gedanken experiment proposed by Salecker and Wigner and consider in detail the physical conditions that permit the emission (absorption) of a photon from (by) the probe particle in canonical quantum mechanics. 
We may then consider the modified conditions induced by the DE-UP.

Classically, a particle of finite extension cannot spontaneously emit another without reducing its internal or kinetic energy \cite{Landau1976}. 
In canonical QM, a non-composite particle does not have internal (i.e. binding) energy, but the wave function of its CoM corresponds to a superposition of position or, equivalently, momentum states. 
Thus, a given positional uncertainty $\Delta x$ corresponds to a momentum uncertainty $\Delta p$, and therefore to an uncertainty in the kinetic energy of order $\Delta E \simeq (\Delta p)^2/2m$. 
This allows the spontaneous emission of additional particles -- for example, the emission of photons from electrons -- without violating conservation of energy or momentum. 
With this is mind, we now reconsider Salecker and Wigner's thought experiment under two different sets of conditions. 
In the first, the particle `tries', and succeeds, in emitting a photon with wavelength $\lambda > \lambda_{\rm C}$. 
In the second, it `tries', and {\it fails}, to emit a photon with $\lambda < \lambda_{\rm C}$.

Prior to the act of measurement, either by an external detector that absorbs it, or via its reabsorption by the particle after reflection at a mirror placed at a distance $r$, the photon is in a superposition of states such that $\Delta \lambda \simeq \hbar/\Delta p \simeq \hbar/\sqrt{2m\Delta E}$. 
The emission of its wave packet takes a time $\Delta t \simeq \Delta \lambda/c \simeq \hbar/(c\sqrt{2m\Delta E})$. 
Thus, if $\Delta E \lesssim mc^2$, then $\Delta \lambda \gtrsim \lambda_C$: the photon wave packet is larger than the particle's Compton wavelength and may escape to communicate with the outside world. 
Specifically, it may traverse a distance $2r$, where $r > \Delta \lambda > \lambda_{\rm C}$, reflect off a mirror and be reabsorbed, yielding a measurement of $r$.

Clearly, if $r < \lambda_C$, the `mirror' cannot lie outside the wave packet of the massive particle and the act of `measurement' involves a self-interaction, in which the particle emits a photon and reabsorbs it within a time $\Delta t \simeq r/c$. This is {\it inevitable} if $\Delta \lambda \simeq c\Delta t < \lambda_{\rm C}$, since the wave packet of the photon will not have sufficient spatial extension, or have travelled far enough over the time-interval $\Delta t$, to escape to the outside world. 
Thus, for $\Delta \lambda \lesssim \lambda_{\rm C}$ ($\Delta E \gtrsim mc^2$), the would-be emitted photon wave packet is `trapped' within the Compton radius of the particle and the associated photon remains {\it virtual}.

Strictly, at this point, the conceptual apparatus of canonical quantum mechanics breaks down and we must switch to the Feynman diagram interactions predicted by QFT. In this picture, the particle emits (and reabsorbs) a virtual photon of wavelength $\lambda$ over a time-scale $t \simeq \lambda/c$. 
The photon is never made {\it real} as this would give rise to a measureable wavelength $\lambda \lesssim \lambda_C$, or, equivalently, $E \gtrsim mc^2$, which is above the threshold for pair-producing particles of mass $m$. 
Nonetheless, in the canonical QM picture this result may be obtained from Salecker and Wigner's bound by setting $r \simeq \Delta \lambda$ in Eq. (\ref{Wiger-Salecker-1}), giving
\begin{eqnarray} \label{}
(\Delta x_{\rm canon.})_{\rm min}(\Delta\lambda) &\gtrsim& \sqrt{\lambda_{\rm C} \Delta\lambda} \gtrsim \Delta\lambda 
\nonumber\\
&\iff& \Delta\lambda \lesssim \lambda_{\rm C} \, .
\end{eqnarray}
To obtain the equivalent bound in the non-canonical theory, represented by Eq. (\ref{DE-UP-2}), we set
\begin{eqnarray} \label{}
(\Delta x_{\rm total})(\Delta\lambda) \gtrsim \sqrt{\lambda_{\rm C}r} + \frac{l_{\rm Pl}^2l_{\rm dS}}{\lambda_{\rm C}r} \gtrsim \Delta\lambda \, .
\end{eqnarray}
This yields $\Delta\lambda \lesssim r_{\rm crit}$, where $r_{\rm crit} \simeq (\Delta x_{\rm total})_{\rm min} \simeq (l_{\rm Pl}^2l_{\rm dS})^{1/3}$ for $m \lesssim m_{\rm crit} \simeq (m_{\rm Pl}^2m_{\rm dS})^{1/3}$ (\ref{critical_mass}). 
Generally, for $m \simeq \alpha_{\rm Q}(m_{\rm Pl}^2m_{\rm dS})^{1/3}$, we have  
\begin{eqnarray} \label{}
\Delta\lambda \lesssim \alpha_{\rm Q}(\Delta x_{\rm total})_{\rm min} \simeq \alpha_{\rm Q}^2\lambda_{\rm C} \, ,
\end{eqnarray}
which automatically ensures $\Delta\lambda \lesssim \lambda_{\rm C}$ for $\alpha_{\rm Q} \leq 1$.

To summarize: In canonical quantum mechanics, photon wave packets with $\Delta\lambda \lesssim \lambda_{\rm C}$ remain `trapped' within the massive particle wave packet, whose minimum extent is given by $(\Delta x_{\rm canon.})_{\rm min} \simeq \sqrt{\lambda_{\rm C}r} \gtrsim \lambda_{\rm C}$. 
In the non-canonical, dark-energy modified theory, the minimum spatial extent of the wave packet and the Compton wavelength of the particle no longer coincide. 
Instead, $(\Delta x_{\rm canon.})_{\rm min} \simeq \alpha_{\rm Q}\lambda_{\rm C}$, which is identified with the classical particle radius, $R$. Photon wave packets with $\Delta\lambda \lesssim \lambda_{\rm C}$ still remain trapped, but only those with $\Delta\lambda \simeq \alpha_{\rm Q}^2\lambda_{\rm C}$ also minimize the positional uncertainty of the CoM.

This suggests that, in the QFT picture, gravitationally-induced modifications of the Feynman diagram structure should yield an expansion in which the main contribution to the particle's self-energy comes the emission and reabsorption of virtual photons with a specific wavelength, $\lambda \simeq \alpha_{\rm Q}^2\lambda_{\rm C}$, In other words, self-interactions with photons of this wavelength should have maximum amplitude, or `weight' in the path integral approach.

Hence, we argue that it {\it is} physically meaningful to consider length-scales $r <  (\Delta x_{\rm total}(r))_{\rm min} < \lambda_{\rm C}$ in dark energy-modified quantum mechanics. 
Though interactions between the particle and its surroundings are possible only for $r \gtrsim \lambda_{\rm C} \gtrsim \Delta x_{\rm total}(r)$, self-interaction is possible within the contiguous regions $\alpha_{\rm Q}^2\lambda_{\rm C} \lesssim r \lesssim \alpha_{\rm Q}\lambda_{\rm C}$ and $\alpha_{\rm Q}\lambda_{\rm C} \lesssim r \lesssim \lambda_{\rm C}$. 
Interestingly, the boundary between the two, $r \simeq (\Delta x_{\rm total})_{\rm min} \simeq \alpha_{\rm Q}\lambda_{\rm C}$, marks the length-scale at which renormalization becomes important for charged particles in QED \cite{Peskin:1995ev,Berestetsky:1982aq}, and our na{\"i}ve picture correctly reproduces a phenomenologically significant length-scale from the relativistic, but non-gravitational, quantum theory of charged particles. 
We may therefore conjecture that, in a more complete theory, including relativistic quantum effects from both dark energy and canonical gravity, the length-scale $r_{\rm min} \simeq \alpha_{\rm Q}(\Delta x_{\rm total})_{\rm min} \simeq \alpha_{\rm Q}^2\lambda_{\rm C}$ should naturally emerge as an {\it effective} cut-off, which minimizes the self-interaction energy of charged particles due to the irremovable `haziness' of the space-time in their vicinity. 



\end{document}